\documentclass[fleqn,usenatbib]{mnras}

\usepackage{newtxtext,newtxmath}
\usepackage{arydshln}
\usepackage[T1]{fontenc}

\DeclareRobustCommand{\VAN}[3]{#2}
\let\VANthebibliography\thebibliography
\def\thebibliography{\DeclareRobustCommand{\VAN}[3]{##3}\VANthebibliography}

\usepackage{graphicx}	% Including figure files
\usepackage{amsmath,bm}	% Advanced maths commands
\usepackage[dvipsnames]{xcolor}
\usepackage{hyperref}
\usepackage{newtxtext,newtxmath}
\usepackage[T1]{fontenc}
\usepackage{adjustbox}
\usepackage{upgreek}
\usepackage{ulem,cancel}%for editing, to be removed before submission
\usepackage{nicefrac,xfrac}
\usepackage{academicons}
\usepackage{tikz}
\newcommand{\RVOASA}{{\sf{O25}}}
\newcommand{\RSOBSA}{{\sf{O60}}}

\newcommand{\RSOBSB}{{\sf{O60q0.7}}}
\newcommand{\RSOBSC}{{\sf{O60q0.5}}}
\newcommand{\RSOBSD}{{\sf{O60q0.3}}}
\newcommand{\CM}{{\sf{O60q0.3cr}}}
\newcommand{\RSOBSE}{{\sf{O60q0.1}}}
\newcommand{\RXOASA}{{\sf{H25}}}
\newcommand{\RAOASA}{{\sf{G25}}}
\newcommand{\RAOASACM}{{\sf{G25cr}}}
\newcommand{\RBOASACM}{{\sf{G50cr}}}

\definecolor{midblue}{rgb}{0.0,0.4,0.7}
\definecolor{mypurple}{rgb}{0.7,0.3,0.8}

\definecolor{PineGreen}{HTML}{008B72}
\definecolor{Berry}{HTML}{FF2052}
\newcommand\aver[1]{\langle#1\rangle}%partial derivative

\newcommand{\dd}{\mathrm{d}}        %differential
\newcommand\sound{_\text{s}} %subscript sound
\newcommand\D{_\text{D}} %subscript dynamo
\newcommand\deriv[2]{\frac{\partial#1}{\partial#2}}%partial derivative
\newcommand\dderiv[2]{\frac{\text{D}#1}{\text{D}#2}}%Lagrangian derivative
\newcommand\mean[1]{\langle #1\rangle}% ensemble average
\newcommand\meanh[1]{{\langle #1\rangle}_{xy}}% horizontal average
\renewcommand\vec[1]{\bm{#1}}% vector convention
\newcommand\kin{_\text{k}} %subscript kinetic
\newcommand\magn{_\text{m}} %subscript magnetic
\newcommand{\BB}{\vec{B}} %Bold B
\newcommand{\U}{\vec{u}} %Bold U
\renewcommand{\mathbfss}[1]{{{\mbox{\boldmath{$#1$}}}}}

\DeclareMathAlphabet{\mathsc}{OT1}{cmr}{m}{sc}
\def\testbx{bx}%
\DeclareRobustCommand{\ion}[2]{%
\relax\ifmmode
\ifx\testbx\f@series
{\mathbf{#1\,\mathsc{#2}}}\else
{\mathrm{#1\,\mathsc{#2}}}\fi
\else\textup{#1\,{\mdseries\textsc{#2}}}%
\fi}

\newcommand{\cm}{\,{\rm cm}}    %cm
\newcommand{\km}{\,{\rm km}}    %km
\newcommand{\p}{\,{\rm pc}}     %parsec
\newcommand{\kpc}{\,{\rm kpc}}  %kpc
\newcommand{\s}{\,{\rm s}}      %seconds
\newcommand{\Myr}{\,{\rm Myr}} %Megayears
\newcommand{\Gyr}{\,{\rm Gyr}}  %Gigayears
\newcommand{\kms}{\km\s^{-1}}    %km/s
\newcommand{\mkG}{\,\upmu{\rm G}} %microGauss
\newcommand{\erg}{\,{\rm erg}}  %erg
\newcommand{\K}{\,{\rm K}}      %Kelvin
\title[Non-linear magnetic buoyancy and galactic dynamos]{Non-linear magnetic buoyancy instability and galactic dynamos}

\author[Y.~Qazi et al.]{Yasin Qazi,$^{1}$\thanks{E-mails: Y.Qazi@newcastle.ac.uk (YQ), anvar.shukurov@ncl.ac.uk (AS), devika.tharakkal@helsinki.fi (DT), frederick.gent@aalto.fi (FAG), abhijit.bendre@epfl.ch (ABB)}\href{https://orcid.org/0009-0008-9513-8761}{\includegraphics[scale=0.5]{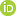}}
A.~Shukurov,$^{1}$\href{https://orcid.org/0000-0001-6200-4304}{\includegraphics[scale=0.5]{orcid_16x16.jpeg}}
D.~Tharakkal,$^{1,2}$\href{https://orcid.org/0000-0002-4563-2277}{\includegraphics[scale=0.5]{orcid_16x16.jpeg}}
F.~A.~Gent$^{3,4,1}$\href{https://orcid.org/0000-0002-1331-2260}{\includegraphics[scale=0.5]{orcid_16x16.jpeg}}
\&
A.~B.~Bendre$^{5,6}$\href{https://orcid.org/0000-0001-5208-8989}{\includegraphics[scale=0.5]{orcid_16x16.jpeg}}
\\
$^{1}$School of Mathematics, Statistics and Physics, Newcastle University, Newcastle upon Tyne, NE1 7RU, UK\\
$^{2}$Department of Physics, University of Helsinki, PO Box 64, FI-00014, Helsinki, Finland\\
$^{3}$Nordita, KTH Royal Institute of Technology and Stockholm University, Hannes Alfv\'ens v\"ag 12, Stockholm, SE-106, Sweden\\
$^{4}$HPCLab, Department of Computer Science, Aalto University, PO Box 15400, FI-00076, Espoo, Finland\\
$^{5}$Department of Physics, Dr. Vishwanath Karad MIT World Peace University, Pune-411038, Maharashtra, India\\
$^{6}$Scuola Normale Superiori di Pisa, Piazza dei Cavalieri, 7, 56126 Pisa, Italy
}

\date{Accepted XXX. Received YYY; in original form ZZZ}

\pubyear{2024}

\begin{document}
\label{firstpage}
\pagerange{\pageref{firstpage}--\pageref{lastpage}}
\maketitle

\begin{abstract}
Magnetic buoyancy (MBI) and Parker instabilities are strong, generic
instabilities expected to occur in most astrophysical systems with sufficiently
strong magnetic fields. In galactic and accretion discs, large-scale magnetic
fields are thought to arise from mean-field dynamo action, particularly the
$\alpha^2\Omega$-dynamo. Using non-ideal MHD equations, we model a section of
the galactic disc where the large-scale magnetic field is generated by an
imposed $\alpha$-effect and differential rotation.  We extend our previous
study of the interplay between magnetic buoyancy and the mean-field dynamo by
incorporating differential rotation, which enhances the dynamo, and cosmic
rays, which amplify magnetic buoyancy.We construct a simple 1D model which replicates
all significant features of the 3D simulations.  Simulations confirm
that magnetic buoyancy can lead to oscillatory magnetic fields and reveal that
it can change the magnetic field parity between quadrupolar and dipolar states.
Differential rotation facilitates this switch in parity, suggesting that the
large-scale magnetic field can adopt a dipolar parity within a few kiloparsecs
of the galactic centre. In contrast, quadrupolar parity may remain predominant in the outer
parts of a galactic disc. Cosmic rays accelerate both the dynamo and the MBI,
supporting oscillatory non-linear states and a spatial magnetic field structure
similar to the alternating magnetic field directions observed in {the haloes of} some edge-on galaxies.
\end{abstract}


%%%%%%%%%%%%%%%%% BODY OF PAPER %%%%%%%%%%%%%%%%%%

\section{Introduction}

The magnetic buoyancy instability (MBI) \citep{N61}, or the magnetic
Rayleigh-Taylor instability is a fundamental process that affects magnetic
fields in stratified plasmas. It develops wherever the strength of a magnetic
field decreases sufficiently rapidly against the gravitational acceleration.
Typical situations where this can arise are in the thin magnetised plasma layer
of galactic \citep{Rodrigues2016,KNPS19,SBADMN19} and accretion discs
\citep{VB97,BH98,Blackman12,JSD14}.  Under the hydrostatic equilibrium, both
magnetic field strength and gas density usually decrease with distance from the
midplane. Since the magnetic field has pressure but not weight, the gas density
is reduced near the midplane where the magnetic field is stronger, producing an
unstable structure.  The interstellar medium of spiral galaxies also contains
cosmic rays which have negligible weight but exert a dynamically significant
pressure. The MBI enhanced by cosmic rays is known as the Parker instability
\citep{Parker1979}.

This ubiquitous instability has a time scale (of the order
of the sound or Alfv\'en crossing time based on the density scale height) much
shorter than the lifetimes of the astrophysical objects, and it must be in its
non-linear state in virtually any object prone to it. The linear stages of both
instabilities are well understood and their dispersion relations have been
obtained for a variety of physical models \citep[e.g.,][see also \citealt{SS21}
and references therein]{Giz1993,FT94,FT95,Kim1997,Rodrigues2016,DT2022a}.  The
non-linear, quasi-stationary states of the MBI and Parker instability are much
less understood, in particular, because they require numerical simulations.

\citet{DT2022a,DT2022b} investigated them in the case of an imposed planar,
unidirectional magnetic field. In a non-rotating system, the instability leads
to a state with large scale heights of both magnetic field and cosmic rays, the
gas layer is correspondingly thin as it is supported solely by the thermal
pressure gradient (and turbulent pressure if available) \citep{DT2022a}.
Rotation changes the non-linear state significantly because gas motions driven
by the instability become helical and can act as a mean-field dynamo
\citep[e.g.,][see also \citealt{HL1997a} and \citealt{1999MOSS} and references
therein]{DT2022b}. As a result, even in the presence of imposed magnetic field,
the magnetic field near the midplane changes profoundly and can reverse its
direction in what appears to be a non-linear, long-period oscillation. Similar
magnetic field reversals occur in the simulations of \citet{JoLe08},
\citet{GJL12} and \citet{MNKASM2013}.

Large-scale magnetic fields in galaxies and accretion discs are produced by a
mean-field ($\alpha$-effect) dynamo \citep[][and references therein]{SS21}, and
\cite{QSTGB23} explore the non-linear instability of a magnetic field generated
by the imposed $\alpha$-effect rather than introduced directly via initial,
boundary or background conditions. Rotation is neglected in this model to
simplify the interaction of the dynamo and the MBI. Magnetic fields generated
by the $\alpha$-effect are helical, and the Lorentz force drives helical
motions which act as a dynamo even without any explicit rotation. As a result,
the system develops non-linear oscillations {of the magnetic field} similar in
their origin to those observed by \citet{DT2022b} in a rotating system with an
imposed non-helical magnetic field.

Here we extend the model of \citet{QSTGB23} to explore the effects of rotation
and cosmic rays on the MBI. We show that the response of the dynamo action to
the instability is even more profound, and the large-scale magnetic field not
only becomes oscillatory, but it can change its parity from quadrupolar (where
the horizontal magnetic field is symmetric with respect to the midplane) to
dipolar state (where the horizontal field is antisymmetric). In this paper, we
seek to reveal, verify and understand these unexpected features of the
non-linear MBI and Parker instability.

As well as a model at the Solar vicinity of the Galaxy, we present a simulation
with parameters typical of the inner parts of spiral galaxies. Our results are
consistent with the complicated structure of the global galactic magnetic
fields in galactic haloes, with large-scale direction reversals as revealed by
observations of the Faraday rotation \citep[see section~3.4.3 of][for a
review]{IB+24}. We are not aware of other convincing explanations of such
complex magnetic structures in galaxies.  Our results show that a quadrupolar
magnetic field produced by the mean-field dynamo action in a thin disc
\citep{SS21} can be transformed into a dipolar field by the magnetic buoyancy
in a rapidly rotating system. This gives credence to the claims that the global
magnetic field within a few kiloparsecs from the centre of the Milky Way has
the dipolar parity \citep{Han17}.

The numerical model used is explained in Section~\ref{sec:Equations}.  Our
simulation results are reported in Section~\ref{sec:Results} in which we
discuss the evolution of the dynamo and MBI in our solutions in
Section~\ref{sec:Rotation}, the effects of model parameters on the growth rates
in Section~\ref{sec:params} and on the parity of the magnetic field in
Section~\ref{sec:parity}. The effects of cosmic rays are included and discussed
in Section~\ref{sec:CM}.  We also consider viscosity and magnetic diffusivity
similar in magnitude to those produced by the supernova-driven turbulence in
spiral galaxies. In Section~\ref{sec:interp} we seek to interpret the results,
examining the $\alpha$-effect during the each stage of the MBI, {and derive
the} turbulent {transport} coefficients which appear in the {mean}
electromotive force in Section~\ref{sec:emf}. Section~\ref{sec:Summary}
summarizes our results {and conclusions}.

%-----------------------------------------
\section{Model description}\label{sec:Equations}

The model and simulations used here are very similar to those of
\citet{QSTGB23} but now include differential rotation. We model isothermal gas
and magnetic field within a {three-dimensional (3D)} Cartesian box with
$x,y$ and $z$ representing the radial, azimuthal and vertical directions,
respectively. The simulation domain extends $4\kpc$ in each horizontal
direction and $3\kpc$ vertically, centred at the galactic midplane. {Although the models all assume a galactocentric distance $R=8\kpc$, for our parameter sweep we exaggerate the rate of shear to more easily excite the MBI so we can further explore the relationship between the MBI and dynamo. {We also} include cosmic rays and {use} more realistic parameters {typical of spiral galaxies}.} We have
tested computational boxes of various sizes from $0.5\kpc$ to $16\kpc$ to
confirm that we capture all essential features of the system. The grid
resolution is $256\times256\times192$ mesh points with a grid spacing of about
$15.6 \p$ along each dimension. The domain size is larger than the expected
vertical and horizontal scales of the instability, and the resolution is
sufficient to obtain convergent solutions.

Table~\ref{tab:parameter_values} summarizes the {common} parameter values
adopted in this study, while Table~\ref{tab:sims} {lists the parameters used
and some indicative results obtained for each simulation} discussed in this
paper. The ratio of shear to rotation is adopted as $q<1$ in some models,
to enhance the MBI relative to the $\alpha^2\Omega$-dynamo and
thus assist the exploration of the relationship between
the two processes. Models with more relevant galactic parameters are also
included.

%---------------------------------------------------------------------
\subsection{Basic equations}

We solve a system of isothermal non-ideal compressible MHD equations using the
sixth-order in space and third-order in time finite-difference \textsc{Pencil
Code} \citep{brandenburg2002,Pencil-JOSS}.  In the local rotating Cartesian
frame $(x,y,z)$, the governing equations are
%------------------------------------------------------------------------
\begin{align}
    \dderiv{\rho}{t} &= -\rho\nabla \cdot \U +\nabla \cdot(\zeta_D\nabla\rho)\,,
            \label{eq:mass_conservation}\\
    \dderiv{\U}{t} &= -g\hat{\vec{z}} - \frac{\nabla P}{\rho}
        +\frac{(\nabla\times\BB)\times\BB}{4\pi\rho}
        +\frac{\nabla\cdot(2\rho\nu{\mathbfss{\tau}})}{\rho}
        -Su_x\vec{\hat{y}}
        -2\vec{\Omega}\times \vec{u}        \nonumber\\
    \label{eq:momentum}
        &\mbox{}\quad +\nabla\left(\zeta_{\nu}\nabla \cdot \U \right)
        +\nabla\cdot \left(2\rho\nu_6{\mathbfss{\tau}}^{(5)}\right)
        -\dfrac{1}{\rho}\U\nabla\cdot\left(\zeta_D\nabla\rho\right)\,,\\
    \deriv{\vec{A}}{t} &=  \alpha \BB+ \U\times \BB - SA_y\vec{\hat{{x}}} - Sx\frac{\partial \mathbf{A}}{\partial y} -\eta \nabla\times \BB
    \label{eq:induction}
    +\eta_6\nabla^{(6)}\vec{A}\,,
\end{align}
%-------------------------------------------------------------------
for the gas density $\rho$, the velocity $\vec{u}$ of the deviations from the
overall rotational pattern and the magnetic vector potential $\vec{A}$. The
vertical gravitational acceleration is $g$, the total pressure $P$, the
magnetic field $\BB=\nabla\times\vec{A}$  and the local angular velocity
$\vec{\Omega}=(0,0,\Omega)$. The physical viscosity and magnetic diffusivity
are $\nu$ and $\eta$, respectively, and $\alpha$ (see
Section~\ref{sect:modelmf}) contributes the $\alpha$-effect that maintains a
large-scale magnetic field via the mean-field dynamo action. The latter is
introduced because we do not include turbulent motions driven by supernovae
which are responsible for the $\alpha$-effect. We note, however, that the
motions driven by the instability also become helical under the action of the
large-scale shear, and this is fully captured by these simulations.

%------------------------------------------------------------------------
\begin{table}
\caption{
Parameters common to all models.
\label{tab:parameter_values}
}
\centering
\begin{tabular}{llll}
\hline
Quantity                   & Symbol          & Value           &Unit     \\
\hline
Grid spacing            & $\updelta\vec{x}$  & 0.0156          &kpc \\
Sound speed                 & $c\sound$          & 15          &km$\s^{-1}$ \\
Initial gas column density  &$\Sigma$            & $10^{21}$   & cm$^{-2}$\\
Shock-capturing viscosity   & $\nu_\text{shock}$ & $(\updelta x)^2\nabla\cdot\vec{u}$ & kpc km s$^{-1}$\\
Shock-capturing diffusivity & $D_\text{shock}$   & $(\updelta x)^2\nabla\cdot\vec{u}$ & kpc km s$^{-1}$\\
Hyper-diffusivities & $\nu_6,\ \eta_6$  & $10^{-12}$ &kpc$^{5}\kms$\\
\hline
\end{tabular}
\end{table}
%------------------------------------------------------------------------

%------------------------------------------------------------------------
\begin{table*}
    \centering
    \caption{MHD simulation parameters and summary
results. The magnitude of the $\alpha$-effect is $\alpha_0$, turbulent magnetic
diffusivity is $\eta$ and the half-thickness of the dynamo-active layer is $h_\alpha$. Galactic rotation
and rate of shear are $\Omega$ and $S$, respectively, with $S=R\,\dd\Omega/\dd
R$, in which $R$ is the galactocentric radius.  {The ratio $q=-S/\Omega=1$, for a flat rotation curve.}
From these parameters, we derive the dynamo
characteristic numbers $R_{\alpha}$ and $R_{\omega}$, given in equation~\eqref{Ral}, and
the dynamo number $D$.  Summary result $\gamma\D$ is the rate of the
exponential growth of the magnetic field strength during the linear phase of
the dynamo and $\gamma_{u}$ is the corresponding growth rate of the
root-mean-square gas speed, due to the subsequent onset of MBI. The
first two models \RVOASA\ and \RXOASA\ have turbulent viscosity $\nu=0.008\kpc\kms$ matching models in
\citet{QSTGB23}, otherwise $\nu=0.3\kpc\kms$.  Model \RXOASA\ has the
highest $R_\alpha=10$, while models with $R_\alpha=5$ are denoted by {\sf{O}}. Models relevant to observed galactic parameters are denoted by {\sf{G}} and
subscript {\sf{cr}} indicates that cosmic rays are included. The global parity of the magnetic field at late stages of the evolution is specified in the last column.}
%The last three models adopt realistic parameters taken from disk glalaxies. CM appended to the model name indicates cosmic rays are included.}
\begin{tabular}{l|cccccc|ccc|ccc}
\cline{1-13}
 Model    &$\alpha_0$  &$\eta$               &$h_\alpha$&$\Omega$                   &$S$                        &$q$&$R_\alpha$&$R_\omega$&$D$  &$\gamma\D$  &$\gamma_u$   &   Magnetic     \\
          &$\!\kms$&km$\s^{-1}\kpc$&pc      &$\!\kms\kpc^{-1}$&$\!\kms\!\kpc^{-1}$&                           &          &          &      &Gyr$^{-1}$& Gyr$^{-1}$&   parity  \\
\cline{1-13}
 \RVOASA  &0.75        & 0.03                &200       &25                         &   $-25$                   &1  &    5     &   $-33.4$&$-167$  & 6.5        &   12.3      &    Dipolar     \\
 \RXOASA  &1.5         & 0.03                &200       &25                         &   $-25$                   &1  &    10    &   $-33.4$&$-334$  & 9.6        &   12.7      &    Quadrupolar \\
 \RSOBSA  &5           & 0.3                 &300       &60                         &   $-60$                   &1  &    5     &   $-18.0$&$ -90$  & 6.1        &   12.4      &    Quadrupolar \\
 \RSOBSB  &5           & 0.3                 &300       &60                         &   $-42$                   &0.7&    5     &   $-12.6$&$ -63$  & 13.3       &   26.7      &    Dipolar     \\
 \RSOBSC  &5           & 0.3                 &300       &60                         &   $-30$                   &0.5&    5     &   $-9.0$ &$ -45$  & 16.4       &   32.5      &    Dipolar     \\
 \RSOBSD  &5           & 0.3                 &300       &60                         &   $-18$                   &0.3&    5     &   $-5.4$ &$ -27$  & 17.5       &   34.5      &    Dipolar     \\
 \RSOBSE  &5           & 0.3                 &300       &60                         &   $-9$                    &0.1&    5     &   $-1.8$ &$ -9 $  & 14.1       &   28.2      &    Dipolar     \\
 \CM      &5           & 0.3                 &300       &60                         &   $-{18}$                 &0.3&    5     &  $-{5.4}$& $-{27}$& 19.6       &   39.1     &    Dipolar     \\
 \RAOASA  &0.3         & 0.3                 &500       &25                         &   $-25$                   &1  &    0.5   &   $-20.8$&$ -10.4$& 1.2        &   1.0      &    Quadrupolar \\
 \RAOASACM&0.3         & 0.3                 &500       &25                         &   $-25$                   &1  &    0.5   &   $-20.8$&$ -10.4$& 1.4        &   1.6      &    Quadrupolar \\
 \RBOASACM&2.5         & 0.3                 &200       &50                         &   $-50$                   &1  &    1.7   &   $-6.7$ &$ -11.4$& 5.4        &   2.1       &    Quadrupolar \\
\cline{1-13}
\end{tabular}
    \label{tab:sims}
\end{table*}
%---------------------------------------------------------------------

The advective derivative is $\text{D}/\text{D}t= \partial/\partial t +
(\vec{U}+\vec{u})\cdot\nabla$ with $\vec{U} = (0,Sx,0)$ the global shear flow
(differential rotation) in the local Cartesian coordinates.  The shear rate is
$S$ (= $R\,\dd\Omega/\dd R$ in terms of the cylindrical radius $R$); for a flat
rotation curve, $\Omega \propto R^{-1}$ and $S = -\Omega$. We neglect the
vertical gradients of the $\Omega$ and $S$ since the observed magnitude of the
vertical gradient of $\vec{U}$ is of the order of $20 \kms \kpc^{-1}$ (section
10.2.3 of \citealt{SS21}, and references therein), leading to a relatively
small velocity lag of order $30\kms$ at $\lvert z\rvert = 1.5 \kpc$.  We apply an external
gravitational force $g$ (see Section~\ref{sect:modelmf}). The isothermal
gas has the sound speed $c\sound = 15 \kms$, which corresponds to a temperature
of $T \approx 2\times 10^4\K$.

The traceless rate of strain tensor $\uptau$ has the form $\tau_{ij} =
\frac{1}{2}(\partial_j u_i + \partial_i u_j)$ (where $\partial_i$ =
$\partial/\partial x_{i}$ and summation over repeated indices is understood.).
Hyperdiffusion with constant coefficients $\nu_6$ and $\eta_6$ is used to
resolve grid-scale instabilities, with $\tau_{ij}^{(5)} = \frac{1}{2}\left[
\partial^5_i u_j + \partial^4_i(\partial_j u_i) \right] -
\frac{1}{6}\partial^4_i(\delta_{ij}\partial_k u_k)$ and $\nabla^{(6)} A_i =
\partial^3_j\partial^3_j A_i$, where $\partial_i^n = \partial^n/\partial x^n_i$
\citep{ABSG2002,Gent2021}.

The artificial viscosity to resolve shocks is introduced with $\zeta_\nu = \nu_{\rm shock}f_{\rm
shock}$ in equation~\eqref{eq:momentum}, where $f_{\rm shock} \propto \lvert \nabla
\cdot \vec{u}\rvert_{-\rm ve}$, is non-zero only in convergent flows \citep[see,
e.g.,][]{Gent2020}.  Following \citet{Gent2020}, we also include the term with
$\zeta_D=D_{\rm shock}f_{\rm shock}$ in equation~\eqref{eq:mass_conservation}
to ensure the momentum conservation in equation~\eqref{eq:momentum}.

The initial conditions represent a hydrostatic equilibrium aside from the
inclusion of a negligible random magnetic field.  The seed magnetic field
applied comprises Gaussian random noise in the vector potential component $A_z$
with a mean amplitude proportional to $\rho^{1/2}(z)$ and the maximum strength
$10^{-6}\mkG$ at $z=0$, such that $B_z=0$. A random initial magnetic field
leads to shorter transients than a unidirectional initial
field.

%---------------------------------------------------------------------
\subsection{Boundary conditions}

The boundary conditions are periodic for all variables
in the $y$ (azimuthal) direction and sliding-periodic along $x$ (radius) to
allow for the differential rotation. To prevent an artificial inward advection
of the magnetic energy through the top and bottom of the domain at $z=\pm
1.5\kpc$, we impose there the conditions $B_x=B_y = \partial B_z/\partial z=0$.
The boundary conditions for the horizontal velocity are stress-free,
\begin{equation}
    \deriv{u_z}{z} = \deriv{u_y}{z} = 0,\quad \text{at } \lvert z\rvert = 1.5\,.
\end{equation}
To permit vertical gas flow across the boundaries without exciting numerical instabilities, the boundary condition for $u_z$ imposes the boundary outflow speed across the ghost zones outside the domain whereas an inflow speed at the boundary tends smoothly to zero across the ghost zones \citep{Gent_SN_ISM_1}. The density gradient is kept at a constant level at the boundaries, with the scale height intermediate between that of the Lockman layer and the galactic halo,
\begin{equation}
    \deriv{\ln\rho}{z} = \pm \frac{1}{0.9\kpc} \quad \text{at } z = \mp 1.5\kpc\,,
\end{equation}
and we note that the value of the scale height imposed at the boundaries has a negligible effect on the results.

%---------------------------------------------------------------------
\subsection{The implementation of the mean-field dynamo\label{sect:modelmf}}
% The characteristics of the galactic mean-field dynamo are understood to depend
% on steep stratification induced by gravity normal to the plane of the disc,
% turbulent helicity due to random motions induced mainly by supernovae denoted
% as the $\alpha$-effect and large-scale shear driven by differential rotation of
% the disc, the latter provided in our model by the sliding periodic boundary
% condition.

%-----------------------------------------------------------------
\begin{figure*}
    \centering
    \includegraphics[width=0.8\textwidth]{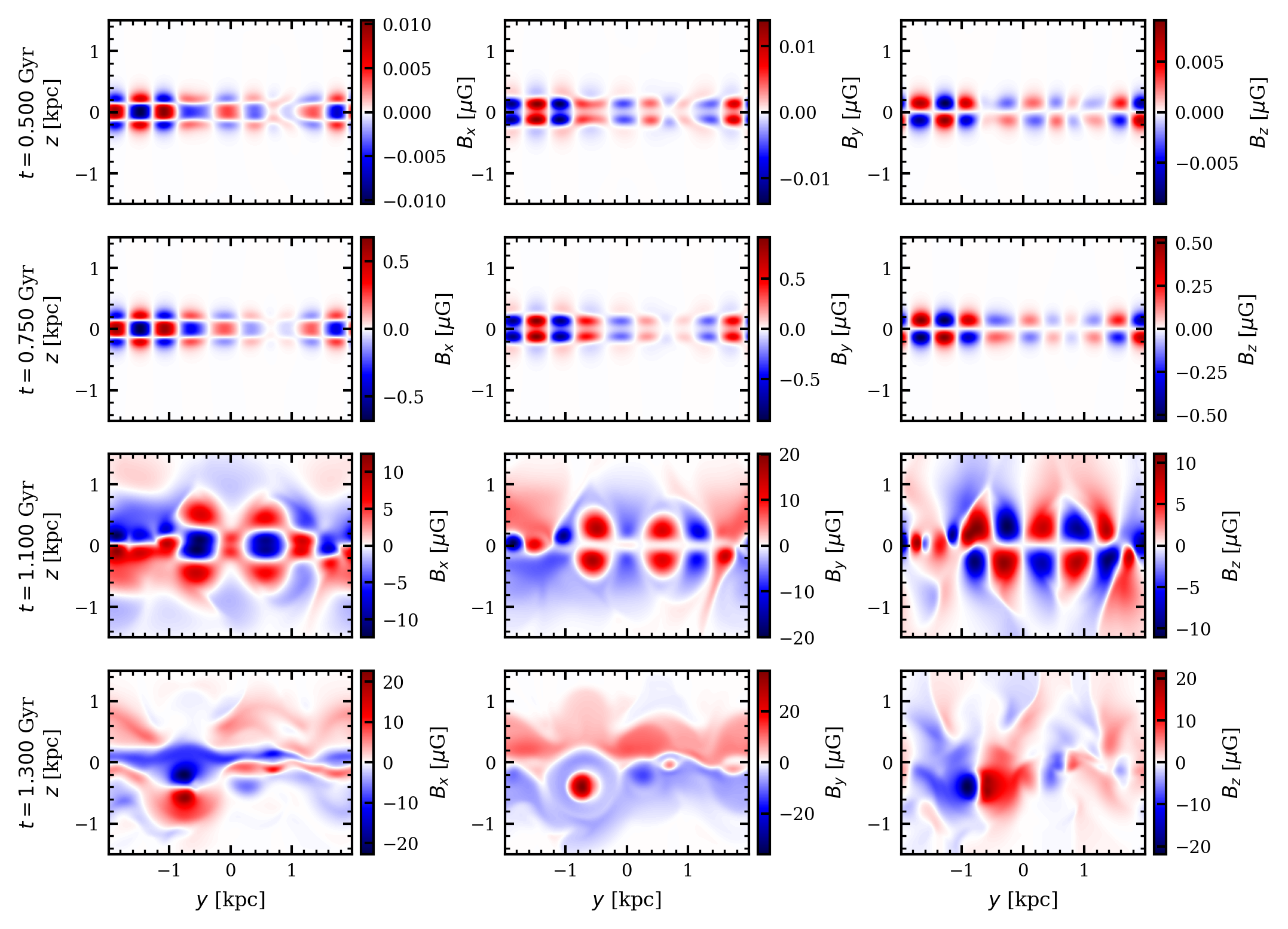}
    \caption{The horizontally averaged magnetic field components $\langle B_x\rangle_{xy}$, $\langle
B_y\rangle_{xy}$ and $\langle B_z\rangle_{xy}$ (columns from left to right) in
the (y,z)-plane at various evolutionary stages in Model \RSOBSD.  During the
{linear phase of the dynamo (upper row, $t = 0.5\Gyr$)} the strength of the magnetic
field grows, while its spatial structure remains largely unchanged (second row,
$t=0.75\Gyr$){, but precipitates the onset of MBI} which  marks the
appearance of large-scale magnetic structure in the magnetic field {late in
the linear phase of the MBI} (third row, $t = 1.1\Gyr$). {The non-linear
phase of the MBI saturates with magnetic structures spanning $\geq1\kpc$}
(lower row, $t=1.3\Gyr$).  }
    \label{fig:field_evolution}
\end{figure*}
%-----------------------------------------------------------------

We adopt a model for the gravitational field appropriate for the Milky Way,
which includes the contribution from the dark matter halo and takes into
account the radial disc mass distribution via the rotation and shear rates.
Following \citet{Ferriere_1998}, we use the gravitational acceleration of
\citet{K&G1989MNRAS} scaled to account for the radial variation of the
gravitational potential,
\begin{equation}
    g =-a_1\frac{{z}}{\sqrt{{z_1}^2+z^2}}\exp{\left(\dfrac{{R_{\odot}-R}}{a_3}\right)} -a_2{\frac{z}{z_2}}\dfrac{R^2_{\odot}+z_3^2}{R^2+z_3^2}
    - 2\Omega(\Omega+S)z\,,
    \label{eq:acceleration}
\end{equation}
where $R_\odot=8.5\kpc$ is the radius of the Solar orbit, $a_1 = 4.4 \times
10^{-14}\km\s^{-2}$ (accounting for the stellar disc), $a_2= 1.7\times 10^{-14}
\km\s^{-2}$ (accounting for the dark matter halo), $z_1=200\p$, $z_2=1\kpc$,
$z_3 = 2.2\kpc$ and $a_3 = 4.9\kpc$.  Stronger gravity at smaller $R$ leads to
a thinner gas disc in the initial state and correspondingly smaller values of
$h_\alpha$ defined below. The Milky Way rotation curve of \citet{1985Clemens}
is used in models for the inner parts of the galactic disc.

Although we aim to explore the interaction of the mean-field (turbulent) dynamo
with the MBI and Parker instability, we do not simulate interstellar turbulence
to ease the control and transparency of the model.In the
absence of turbulence driven by supernovae, radiative pressure and
self-gravity, we impose an $\alpha$-effect , which
represents the summation of these turbulent processes on the mean-field dynamo
action with parameters typical of spiral galaxies. We use the same form of the $\alpha$-effect as
\cite{QSTGB23}, but which in this case is also now enhanced by the
effect of the galactic shear. The $\alpha$-effect is antisymmetric in $z$,
localized around the midplane within a layer of $2h_{\alpha}$ in thickness and
smoothly vanishing at larger altitudes,
\begin{equation}
    \label{eq:alpha}
    \alpha(z)=\alpha_0
    \begin{cases}
    \displaystyle
    \sin \left(\pi z/h_\alpha\right)\,, &\lvert z\rvert \leq h_\alpha/2\,,\\
    \displaystyle
    (z/\lvert z\rvert) \exp \left[-\left(2z/h_\alpha-z/\lvert z\rvert\right)^2\right]\,, &\lvert z\rvert>h_\alpha/2\,.
    \end{cases}
\end{equation}
Taking the curl of equation~\eqref{eq:induction} this applies symmetric
amplification to $\vec{B}$ with a dependence on $\cos z$, with strength
$\alpha_0$ maximum at the midplane $z=0$, and antisymmetic amplification of
$\partial \vec{B}/\partial z$. The vertical extent of the dynamo-active layer
is $h_{\alpha}$ on each side of the midplane. The smaller $h_{\alpha}$, the
stronger the vertical gradient of the magnetic field and the more it is
buoyant. In Sections~\ref{sec:Rotation}--\ref{sec:emf}, we explore generic
features of the MBI and adopt $h_{\alpha} = 0.3 \kpc$ (equal to the initial
density scale height) to make the instability stronger, this also allows for a
more direct comparison with the case of a non-rotating system \citep{QSTGB23}.

As listed in Table~\ref{tab:sims}, we include several models which{{, while still assuming a galactic radius of 8.5~kpc,}}
explore extreme values for $R_{\alpha}$ and $R_\omega$ in order to {better} discern how
rotation affects the non-linear evolutionary phase of the system. {The models {\RAOASA,
\RAOASACM} and {\RBOASACM} consider different galactocentric distances}.
{\RBOASACM\ uses parameters which match M31} at $R=3\kpc$. We adopt the
magnitude of the $\alpha$-effect $\alpha_0 = 2.25 \kms$ (e.g., p.317 of
\citealt{SS21}).

The dynamo intensity (both the rate of exponential growth of the magnetic field
strength at an early stage and its steady-state magnitude) depends on the
dimensionless parameters
\begin{equation}\label{Ral}
R_\alpha=\alpha_0 h_\alpha/\eta\quad \text{and} \quad   R_\omega=Sh_\alpha^2/\eta\,,
\end{equation}
which quantify the magnetic induction by the $\alpha$-effect and differential
rotation, respectively. When $R_{\alpha} \ll \lvert R_\omega\rvert $, the magnetic field is
mostly sensitive to their product \citep[][section 11.2]{SS21} known as the dynamo number,
\begin{equation}\label{DynNum}
     D=R_\alpha R_\omega\,.
\end{equation}

\citet{QSTGB23} considered a non-rotating system with an imposed
$\alpha$-effect, a form of the mean-field dynamo known as the
$\alpha^2$-dynamo. Here we include differential rotation to obtain a stronger
magnetic field amplification mechanism, the $\alpha^2\Omega$-dynamo.

\begin{figure*}
    \centering
    \includegraphics[trim=0.0cm 0.0cm -23.9cm 0.0cm,clip=true,width=1\textwidth]{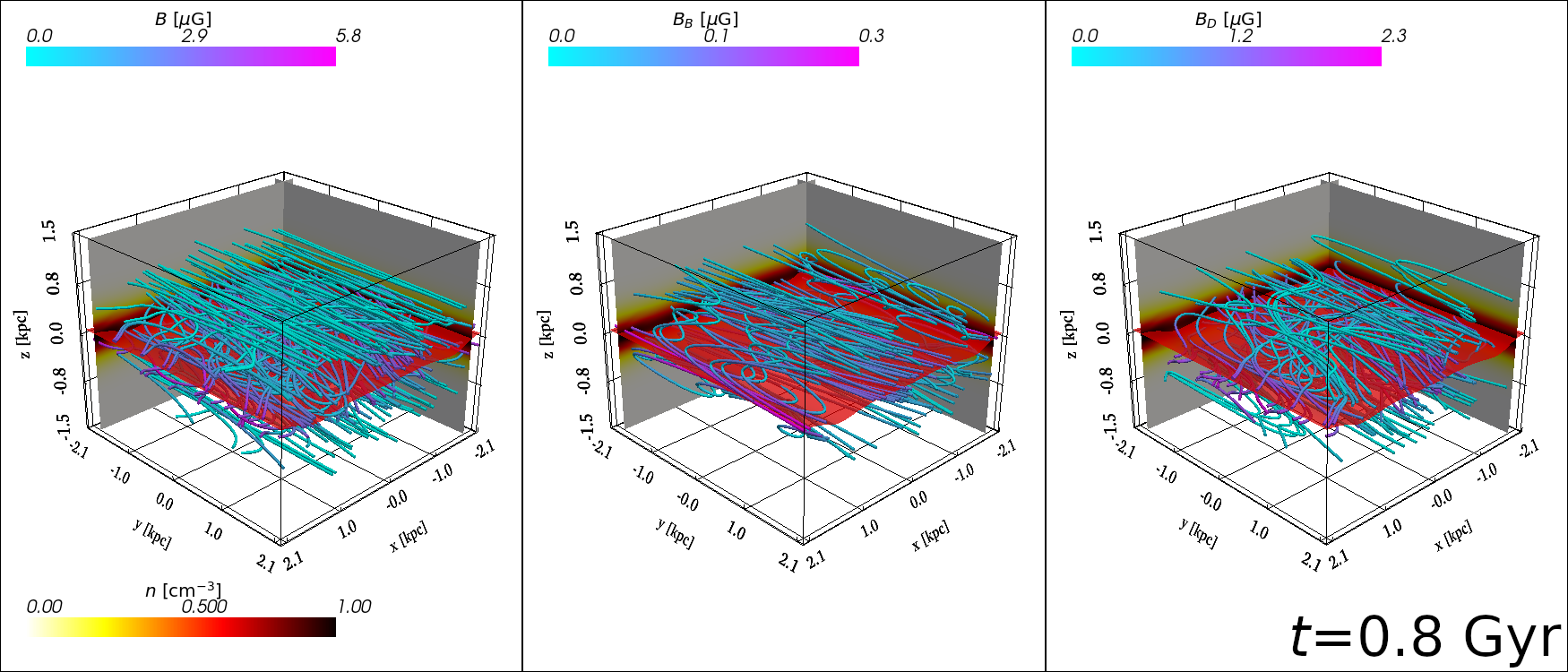}\\
    \includegraphics[width=0.72\textwidth]{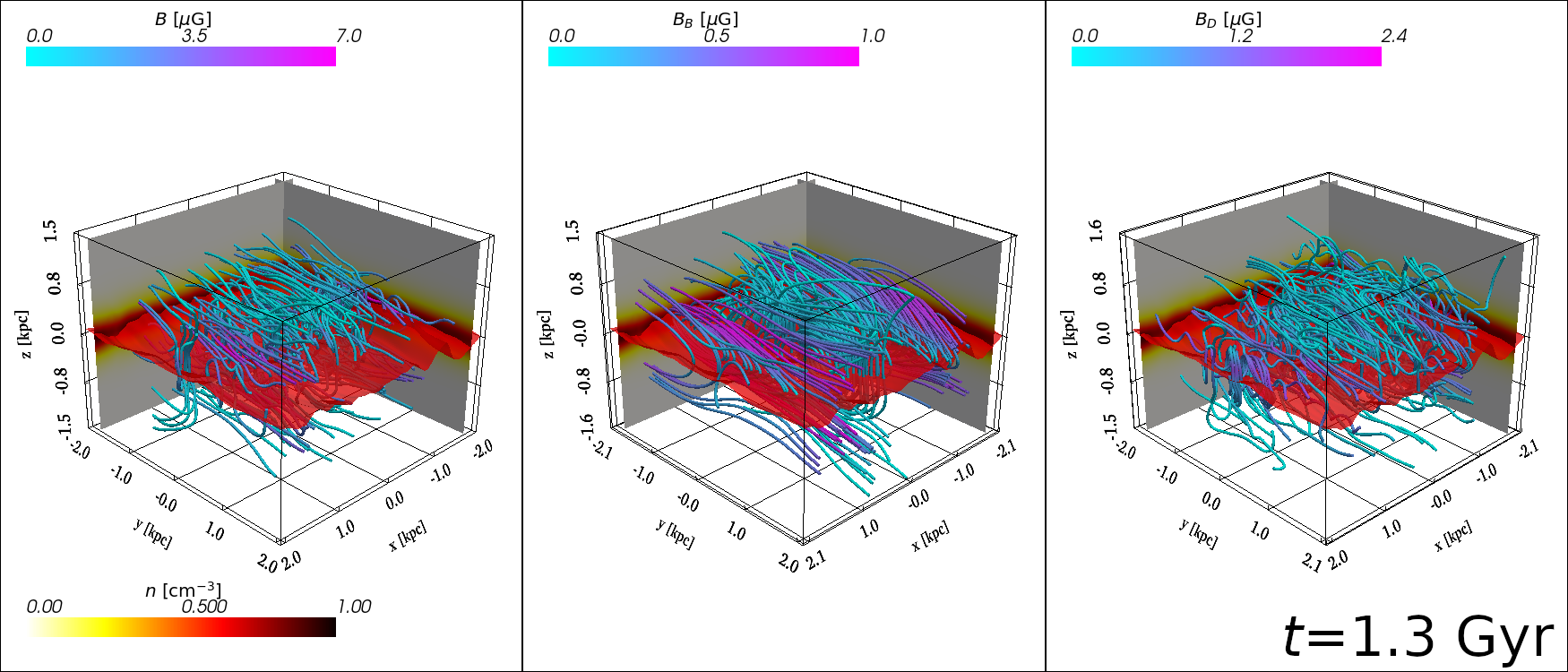}
    \includegraphics[trim=0.0cm -2.0cm 0.0cm 0.0cm,clip=true,width=0.27\textwidth]{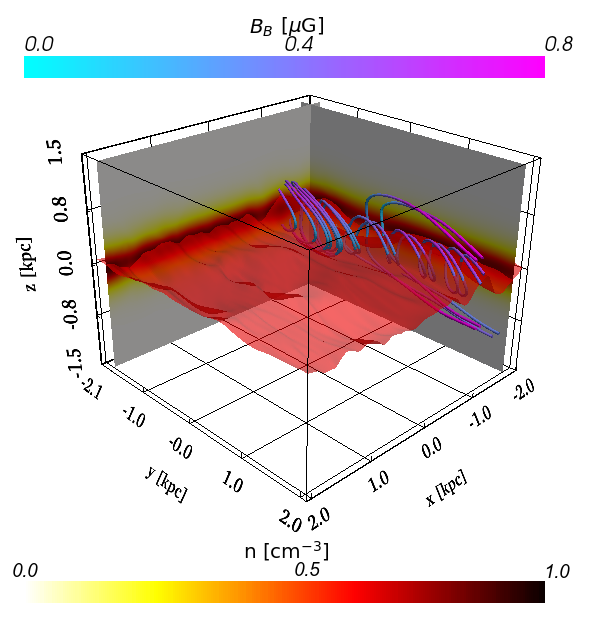}
    \begin{picture}(1.0,0)
     \put(-410.07,300.350){\sf{\bf{(a)}}}
     \put(-290.07,300.925){\sf{\bf{(b)}}}
     \put(-170.07,300.500){\sf{\bf{(c)}}}
     \put(-410.07,140.350){\sf{\bf{(d)}}}
     \put(-290.07,140.925){\sf{\bf{(e)}}}
     \put(-170.07,140.500){\sf{\bf{(f)}}}
     \put(  -0.07,140.500){\sf{\bf{(g)}}}
     \end{picture}
    \caption{The magnetic lines in Model {\RSOBSD} of the total magnetic field
$\BB$ (panels (a) and (d)) which is separated using a Gaussian kernel of the
smoothing length $\ell=200\p$ into contributions characteristic of the magnetic
buoyancy $\BB_\text{B}$ with the larger scales (panels (b) and (e)) and those
of the dynamo $\BB_\text{D}$ at smaller scales (panels (c) and (f) ). The red isosurface represents the gas number density at $0.7\cm^{-3}$.  \label{fig:streamlines} \label{fig:loops} 
{Panel (g) shows a part of $\BB_\text{B}$ taken from  panel (e) where the Parker loops are easily identifiable.}}
\end{figure*}

\section{Results}\label{sec:Results}
Model \RSOBSD\
{(see
Table~\ref{tab:sims}) is used to present our main results whereas other models address details.
Models {\RVOASA} and {\RXOASA} use Prantdl numbers matching those
with $\Omega=0$ of \citet{QSTGB23} in order to isolate the effects of rotation.
The runs \RSOBSA --\RSOBSE\ are used to explore how rotation affects the final
steady-state magnetic field parity, while models {\RAOASA, \RAOASACM\ and \RBOASACM} use
parameters which reflect conditions at various galactocentric radii with rotation curve for the Milky Way of \citet{1985Clemens}.

%---------------------------------------------------------------------
\subsection{The interaction of dynamo and magnetic buoyancy}\label{sec:Rotation}

The main features of the interaction of the dynamo and MBI can be illustrated
using Model \RSOBSD\ in which their growth rates and characteristic scales are
quite different. Since $\lvert R_\omega\rvert \approx R_\alpha$ in this model,
the estimate of the dynamo length scale of about $\lambda=300\p$ obtained by
\citet[][section~3]{QSTGB23} for the $\alpha^2$-dynamo remains a valid
approximation; the wavelength of the MBI is much larger, of order 1--2\,kpc.
{This scale separation is supported by an inspection of the evolving field
structure in Fig.~\ref{fig:field_evolution}.}

%--------------------------------------------------------------------------------------
\begin{figure*}
    \centering
    \includegraphics[width=\textwidth]{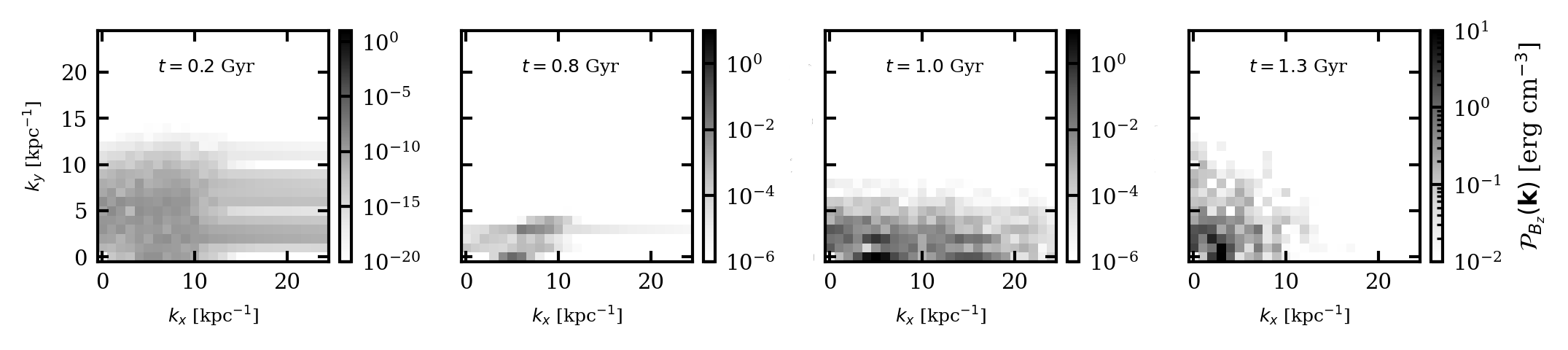}
    \caption{Two-dimensional power spectra in the ($k_x$,$k_y$)-plane of $B_z$
in Model \RSOBSD\ at $z=385\p$ during the evolution of the mean-field dynamo
and onset of the MBI (leftmost and middle panels) through to a stationary
state (right).}
    \label{fig:Power_spec}
\end{figure*}
%--------------------------------------------------------------------------------------

At early times (upper row), magnetic field produced by the dynamo at a
relatively small scale is too weak to be buoyant, but, as its strength
increases, it {becomes susceptible to distortion} by magnetic buoyancy
(second row).  The spatial structure dominated by the MBI is shown in the third
row corresponding to the time when the system enters the stationary state. Here
the magnetic field has spread to large altitudes and the vertical magnetic
field has become locally comparable in magnitude to the horizontal {field}
components. The vertical parity of the magnetic field remains quadrupolar (the
same as in the dynamo field): the horizontal field is symmetric with respect to
the plane $z=0$ while the vertical field is antisymmetric. Despite the strong
difference in the spatial scales, this structure is maintained by the dynamo
action, this is a true symbiosis of the two processes.

The evolution described above is quite similar to that discussed by
\citet{QSTGB23}, where $\Omega=0$ and $S=0$, {where at early times (the
first two rows of Fig.~\ref{fig:field_evolution}) the magnetic field has a
small scale controlled by the dynamo and is confined to the region $z<\lvert
h_{\alpha}\rvert $ where the $\alpha$-effect is imposed and evolves as the
dynamo eigenfunction. The slight variation of the solution along $y$ likely reflects a weak buoyancy of the magnetic field. However the addition of rotation yields} enhanced
regular magnetic patterns at $t\leq1.1\Gyr$, {due to the stronger shear dynamo
action}.  The first three rows are useful to compare with fig.~4 of
\citet{QSTGB23} where the evolution is slower and the magnetic structures are
less regular because of a weaker dynamo. {At later times the magnetic field
spreads out of the disc because of the magnetic diffusion and the {strong} signs
of the magnetic buoyancy emerge. This is particularly manifested in the
enhancement of $B_z$ via the stretching of $B_x$ and $B_y$ by the vertical velocity dependent on $x$ and $y$.} However, thus far rotation has not introduced any
qualitative changes into the system.

This difference {in the characteristic scales motivates} us to separate the two
types of the magnetic field using the Gaussian smoothing \citep{Gent2013}.  The
buoyancy-driven part is obtained from the total magnetic field
$\vec{B}(\vec{x},t)$ as
\begin{equation}\label{GS}
    \BB_\text{B}(\vec{x},t) = \int_{V}B(\vec{x}^\prime,t)\,G_{\ell}(\vec{x}-\vec{x}^\prime)\,\dd^3\vec{x}^\prime\,,
\end{equation}
where the integration extends over the whole domain volume with the smoothing
kernel $G_\ell(\vec{\zeta}) = (2\pi\ell^2)^{-3/2}\exp{[-\lvert
\vec{\zeta}\rvert^2/(2\ell^2)}]$ and $\ell = 200\text{--}300\p$ chosen to be
close to the dynamo scale $h_\alpha$. The remaining part of the magnetic field
$\vec{B}\D = \vec{B}-\vec{B}_\text{B}$ has scales smaller than $\ell$. It is
mostly due to the dynamo action but also contains random fields produced by
non-linear effects at the later stages of the evolution.

Applying this filter, we illustrate in Fig.~\ref{fig:streamlines} the 3D
field structures, including the loops produced by the MBI (which are not very
prominent because the magnetic field is rather disordered even at larger
scales) and the magnetic field generated by the dynamo.  Magnetic field lines
are plotted (left to right) for the total field, $\BB$, the buoyancy-driven
field $\BB_\text{B}$ and the dynamo field $\vec{B}\D$, before the development
of the MBI at $t=1\Gyr$ and after it has saturated at $t=1.3\Gyr$.

The instability produces buoyant loops of a large-scale magnetic field at a
kiloparsec scale. These `Parker loops' are expected to lie largely in the
azimuthal direction, the direction of the large-scale field.  This corresponds
to the `undular' modes (with wavevector parallel to the magnetic field $\BB$),
which are expected to dominate over the `interchange' modes (with wavevector
perpendicular to $\BB$), derived from linear analyses of the instability
\citep[see, e.g.,][]{MTSK1993}. Such twisted loops are seen in panel (g) in
Fig.~\ref{fig:loops} which displays a small portion of the filtered magnetic
field in the non-linear stage of the evolution.

The restructuring of the magnetic field of model \RSOBSD\ by the MBI is
quantified in Fig.~\ref{fig:Power_spec}. This shows the two-dimensional power
spectra of the $z$-component of the magnetic field at times indicated.  These
confirm the evolution pattern visible in Fig.~\ref{fig:field_evolution}. Over
time the dominant horizontal scales $2\pi k_x^{-1}$ and $2\pi k_y^{-1}$ of the
magnetic field grow larger.  At $t\lesssim 0.2 \Gyr$ the energy is confined to
azimuthal scales $k_y\lesssim 10\kpc^{-1}$, {while} the radial scales extend to
$k_x>20\kpc^{-1}$.

{The dominant azimuthal wavenumber $k_y$ of the magnetic field decreases under the influence of rotation. Figure~6 of \citet{QSTGB23},
where $\Omega=0$, shows that $k_y=15\kpc^{-1}$ at $t=1\Gyr$, whereas $k_y=5\kpc^{-1}$
during a similar stage of evolution in  model \RSOBSD\ where $\Omega=60 \kms
\kpc^{-1}$ (see Fig.~3).} The dominant horizontal scales increase through to $t=1.3\Gyr$ to
reach $k_x\approx 4$ and $k_y\approx 2$.  These wavenumbers correspond to
scales of 1--2$\kpc$, characteristic of the MBI. As the peak wavenumbers
decrease further, due to the onset of MBI, the spectrum becomes broader, as the
MBI excites a wider range of unstable modes. {Structures of these scales are visible in Figures 1 and 2.}

%-----------------------------------------------------------------
\begin{figure}
    \centering
    \includegraphics[width=0.45\textwidth]{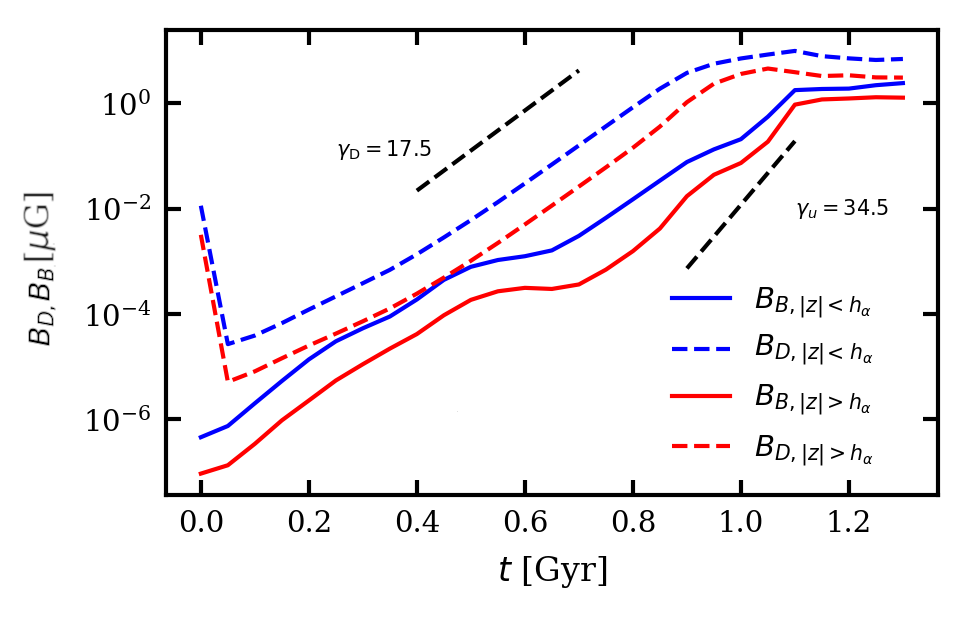}
    \caption{The evolving magnitude of the magnetic field in model \RSOBSD\ at
larger (solid) and smaller (dashed) scales, obtained using the scale separation
of equation~\eqref{GS} with $\ell=300\p$, averaged over $\lvert z \rvert <
h_{\alpha}$ (blue) and $\lvert z \rvert > h_{\alpha}$ (red).  Dashed lines
(black) indicate the exponential growth at the rates presented in
Table~\ref{tab:sims}.
    }
    \label{fig:ts_mag}
\end{figure}
%-----------------------------------------------------------------

To investigate the growth rates of the instabilities, the magnetic field is
separated into $\BB_\text{B}$ and $\BB_\text{D}$ within and without the
distance $h_\alpha$ from the midplane.  As shown in Fig.~\ref{fig:ts_mag},
after the initial transient decay, the total field strength grows in time up to
a stationary state.  The rate of growth $\gamma_\text{D}$ of the total magnetic
field within $\lvert z\rvert\leq h_\alpha$ is estimated for each model in
Table~\ref{tab:sims} during the interval after
which its strength reaches 10 times its minimum through to 5 per cent of its
maximum.  In the case of Model \RSOBSD\ this interval spans $0.2\Gyr\lesssim t
\lesssim0.75\Gyr$ and has $\gamma_\text{D}=17.5\Gyr^{-1}$.

Once the magnetic field becomes buoyant, velocity perturbations start growing
exponentially. In all models the growth rate $\gamma_u$ is
measured between the root-mean-square velocity perturbation
$u_\text{rms}$ exceeding 10 times its minimum and attaining 
10 per cent of its maximum.  In Model \RSOBSD\
(Fig.~\ref{fig:ts_mag}), $\gamma_u=34.5\Gyr^{-1}$ at $0.75\Gyr\lesssim t
\lesssim 1\Gyr$.

As shown in Fig.~\ref{fig:ts_mag}, the magnetic field $\BB_\text{D}$ of 
smaller scale (dashed blue), mainly driven by the dynamo action, has a near
constant growth rate through to the stationary state at $t\gtrsim1.1\Gyr$. The
dynamo action is localized at {$\lvert z\rvert\lesssim h_\alpha$}, but the
magnetic field spreads diffusively to larger altitudes (dashed red) where,
although much weaker, {it has} the same {growth} rate.  At
$t\lesssim0.5\Gyr$, the magnetic field at larger scales $\BB_\text{B}$
(solid lines) represents just the large-scale tail of the leading dynamo
eigenfunction.  However, its behaviour subsequently changes, stagnating for
some 200~Myr before growing further, while $\BB_\text{D}$ continues to grow.

This transition is not observed when differential rotation is absent
\citep{QSTGB23} where, instead, the growth rate of $\BB_\text{B}$ exceeds that
of $\BB_\text{D}$ because the dynamo is weaker.  Here, following the
transition, MBI drives a new dynamo action on $\BB_\text{B}$.  The transitory
stagnation in $\BB_\text{B}$ may be due to the reduction of the radial scale of
the magnetic field by the large-scale velocity shear, which is reflected in the
growth of small scale structure in $k_x$ between 0.8 and 1\,Gyr  (without any
significant change in $k_y$) visible in Fig.~\ref{fig:Power_spec}.

%---------------------------------------------------------------------
\subsection{Effect of parameters on the growth rates.}\label{sec:params}

The structures produced by the $\alpha^2\Omega$-dynamo and the MBI grow
exponentially during their linear stages at different rates, becoming strongly
intertwined during non-linear stages of the instabilities when the Lorentz
force becomes dynamically significant and the system evolves into a stationary
state.  The two processes respond differently to the system parameters.  For
example, reducing {only} $h_\alpha$ makes the dynamo action weaker, because the
dynamo parameters $R_\alpha$ and $R_\omega$ become smaller, but enhances the
MBI, because the gradient of the magnetic field strength increases with
$h_\alpha^{-1}$.  Furthermore, the MBI is sensitive to both the magnetic
diffusivity and the kinematic viscosity whereas the dynamo action is relatively
insensitive to the kinematic viscosity.

The parameters and outcomes presented in Table~\ref{tab:sims} are designed to
aid identification of the physical processes responsible for salient features
of the system's steady state.  Some unrealistic parameter values have been
chosen to enhance the difference in the properties of the dynamo and MBI. We
flag such parameter choices and emphasise results that were obtained for the
parameter values typical of spiral galaxies.

Models \RVOASA\ and \RXOASA\ match parameters in \citet{QSTGB23}, but with the
addition of differential rotation. In
Model \RVOASA, $\gamma_\text{D}$ is boosted by differential
rotation from $1.6\Gyr^{-1}$ to $6.5\Gyr^{-1}$ and $\gamma_u$ from
$2.1\Gyr^{-1}$ to $12.3\Gyr^{-1}$.  In contrast, Model~\RXOASA\
has $\gamma_\text{D}$ reduced from $12.4\Gyr^{-1}$ to
$9.6\Gyr^{-1}$ and $\gamma_u$ from $25.8\Gyr^{-1}$ to $12.7\Gyr^{-1}$. Here,
and in Models \RSOBSA--\RSOBSE, the dynamo-dominated
solutions have growth rates controlled by the  dynamo
number $D$ alone rather than by both $R_\alpha$ and
$R_\omega$. Conditions under which
the $\alpha\Omega$-dynamo prevails over the $\alpha^2\Omega$ mechanism are
obtained by comparing the growth rates of the $\alpha^2$ and $\alpha\Omega$
dynamos. For the $\alpha^2$-dynamo, \citet{SSR83} show that its growth
rate is estimated as $\gamma_{\alpha^2}\simeq R_\alpha^2$
for $R_\alpha\gg1$. Since the growth rate of the $\alpha\Omega$ dynamo is
estimated as $\gamma_{\alpha\Omega}\simeq |D|^{1/2}$ for $|D|\gg1$
\citep[e.g.,][]{Ji14}}, we have
\begin{equation}\label{gamratio}
\dfrac{\gamma_{\alpha^2}}{\gamma_{\alpha\Omega}}
\simeq\dfrac{R_\alpha^2}{R_\alpha^{1/2}\lvert R_\omega\rvert^{1/2}}
\simeq\dfrac{R_\alpha^{3/2}}{\lvert R_\omega\rvert^{1/2}},
\end{equation}
for large $R_\alpha$ and $|R_\omega|$. The condition for the dominance of the
$\alpha\Omega$ mechanism
$\gamma_{\alpha\Omega}>\gamma_{\alpha^2}$ reduces to
\begin{equation}\label{eq:gratio}
\lvert R_\omega \rvert > R_\alpha^3.
\end{equation}

In the case of Model \RVOASA, where $\lvert R_\omega\rvert\gg R_\alpha$, this
is an $\alpha\Omega$-type dynamo. Thus, increasing $\lvert D\rvert$ leads to an
increase in $\gamma_\text{D}$. With the larger $R_\alpha$ of Model \RXOASA, the
solution remains dominated by the $\alpha$-effect since $R^3_\alpha > \lvert D
\rvert$, making it an $\alpha^2$-dominated dynamo, but with the growth impeded
by a competing shearing effect.

In Models \RSOBSA--\RSOBSE, $\lvert R_\omega\rvert$ is sufficiently close to
$R_\alpha$ to make the dynamo relatively insensitive to $D$. The
$\alpha$-effect is dominant and capable of producing large values of
$\gamma_\text{D}$. However, shear in such cases can impede the $\alpha$-effect,
leading to an increase in the growth rate $\gamma_\text{D}$ as $q=-S/\Omega$
decreases from 1 to 0.3. As $S\rightarrow0$, however, the $\alpha^2
\Omega$-dynamo weakens, as evident in Table~\ref{tab:sims} for $q<0.3$. The
velocity growth rate $\gamma_u\simeq2\gamma_\text{D}$ reflects the relative
strength of magnetic buoyancy present.

For the more realistic Solar neighbourhood parameters of Model~\RAOASA, where
$\lvert R_\omega\rvert \gg R_\alpha$ and $D=-10.4$, the solution is sensitive
to $D$. Both $\gamma_\text{D}$ and $\gamma_u$ are approximately $1\Gyr^{-1}$,
appropriately smaller than in \RVOASA, where $D=-167$. In \RBOASACM, $\lvert
R_\omega\rvert$ is not significantly greater than $R_\alpha$, suggesting that
the dynamo is likely dominated by the $\alpha$-effect.

%---------------------------------------------------------------------
\subsection{Magnetic field symmetry}\label{sec:parity}

\begin{figure}%---------------------------------------------------------------------
    \centering
    \includegraphics[width=0.52\textwidth]{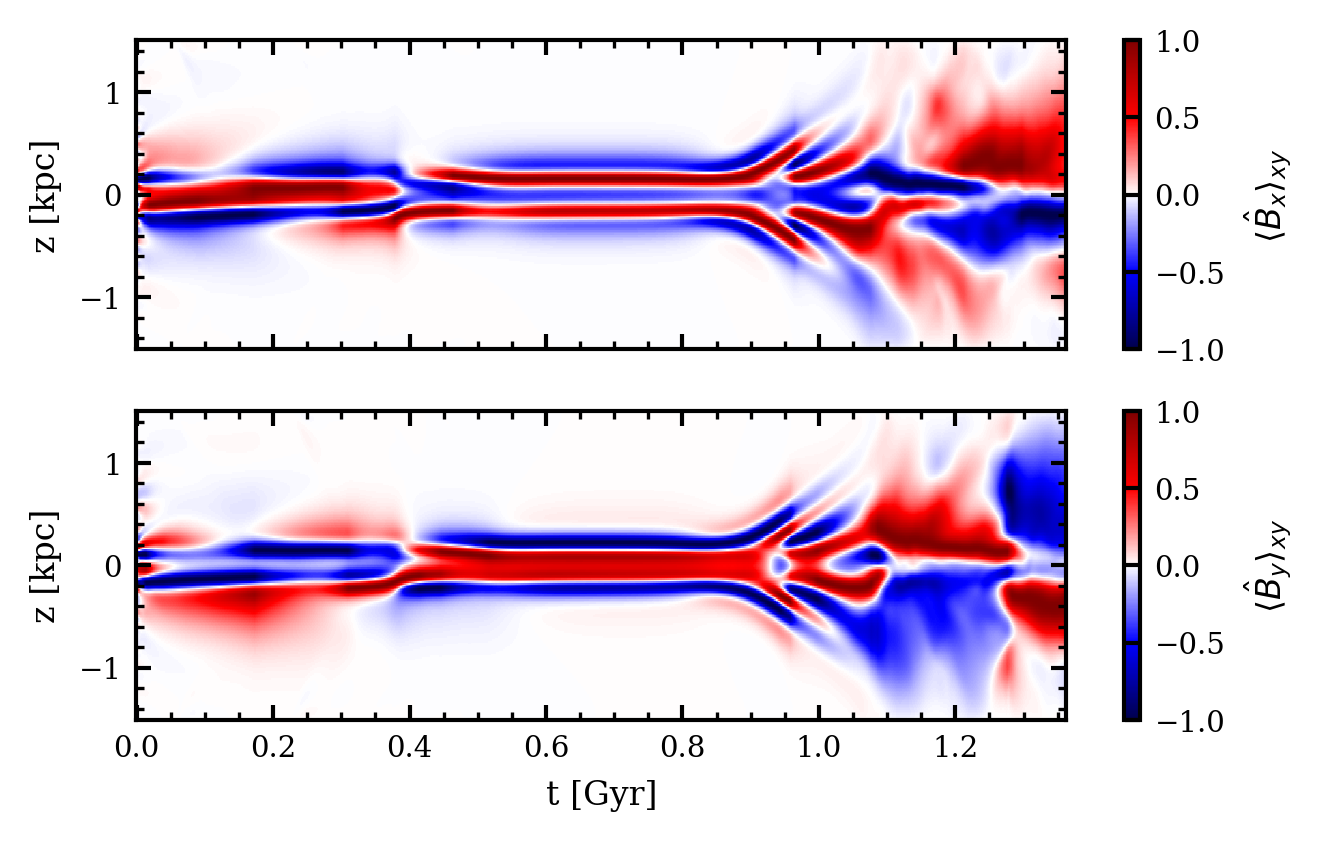}
    \caption{The evolution of the horizontally averaged magnetic field
components $\langle\widehat{B}_x\rangle_{xy}$ (upper panel) and $\langle
\widehat{B}_y \rangle_{xy}$ (lower panel) in Model {\RSOBSD}.  The hat
indicates that each component has been normalized to its maximum magnitude at
each time.}
    \label{fig:xy_averages}
\end{figure}%---------------------------------------------------------------------

A fundamentally new consequence of differential rotation
(which enhances the dynamo action and, indirectly, the MBI) emerges in the
late non-linear stage, where the magnetic field structure changes from a
predominantly quadrupolar to a dipolar symmetry
(see the fourth row of Fig.~\ref{fig:field_evolution} and
Fig.~\ref{fig:xy_averages}).  In a perfectly quadrupolar structure, the
horizontal magnetic field components are symmetric with respect to the midplane
whereas the vertical field is antisymmetric,
\begin{equation}\label{QS}
B_{x,y}|_{z<0} = B_{x,y}|_{z>0}\,,\qquad B_z|_{z<0} = -B_z|_{z>0}\,.
\end{equation}
A dipolar field has the opposite symmetry,
\begin{equation}\label{DS}
B_{x,y}|_{z<0} = -B_{x,y}|_{z>0}\,,\qquad B_z|_{z<0} = B_z|_{z>0}\,.
\end{equation}
Although the symmetry surface is not flat but rather undulates around $z=0$,
the change of the field symmetry is obvious at about $t=1.1\Gyr$ despite the
fact that the imposed $\alpha$-effect, confined to relatively thin layer,
sustains a magnetic field of quadrupolar parity \citep[Section~11.3.1
of][]{SS21}, and the buoyancy does not change that in the early non-linear
stage.

Fig.~\ref{fig:xy_averages} illustrates how the parity of the magnetic field is
transformed as a consequence of the MBI under the effects of rotation.
Throughout the linear stage of the MBI the magnetic field grows monotonically
before changing parity at $t\geq 1.1\Gyr$ when it becomes strong enough to make
the system essentially non-linear. The figure shows the evolution of the
horizontally averaged magnetic field components $\langle B_x\rangle_{xy}$ and
$\langle B_y\rangle_{xy}$ from Model~\RSOBSD, normalized to their maximum
values at each time to better reveal the field structure at early times when it
is still weak.

Models for which the intensity of the MBI, as indicated by $\gamma_u$
(Table~\ref{tab:sims}), is up to twice the intensity of the
$\alpha^2\Omega$-dynamo, as indicated by $\gamma_\text{D}$, appear to support
dipolar magnetic fields in the non-linear steady state. Models where $\gamma_u
\lesssim \gamma_\text{D}$ exhibit quadrupolar structures. A strong MBI is
easier to excite at a reduced scale height $h_\alpha$.

Models {\RVOASA} and {\RSOBSA} appear to counter this trend, with
$\gamma_\text{D}$ and $\gamma_u$ being quite similar between the two models.
Model {\RVOASA} is more sensitive to the dynamo number $D$ and yields a dipolar
field, while Model {\RSOBSA} is more sensitive to $R_\alpha$ and yields a
quadrupolar field. For dynamos in which $\lvert R_\omega\rvert\gg R_\alpha$, it
is therefore easier to excite dipolar modes when the MBI is strong. Models
{\RVOASA} and {\RXOASA} use parameters from the simulations R5h2 and R10h2 as
part of a suite of simulations without rotation in \citet{QSTGB23}. None of
those models exhibit this change in parity.

\citet{MNKASM2013} investigate a system that includes magneto-rotational and
Parker instabilities. Their fig.~6 shows clear, regular time reversals of the
magnetic field similar to those found by \citet{QSTGB23}.  Furthermore, fig.~10
of \citet{MNKASM2013} shows the distribution of RM obtained from their
numerical results, which corresponds to a dipolar magnetic field. However,
simulations of the galactic dynamo in the Solar vicinity of the Milky Way
driven by the supernova-driven turbulence have so far produced quadrupolar
solutions \citep{Gressel2008,OGPhD,Gent2013,GMK23}.

\citet{TSS2009} find that the Faraday rotation measures (RM) of the
extragalactic radio sources suggest a dipolar structure of the horizontal
magnetic field outside the disc of the Milky Way although the vertical magnetic
field has a quadrupolar symmetry.  However, \citet{Mao+10} analyse similar data
to conclude that the vertical magnetic field in the Milky Way halo does not
have a clear-cut symmetry near the Sun, whereas \citet{Mao+12} find that the
toroidal field in the Milky  Way halo is similarly directed on both sides of
the Galactic disc. These authors present a review of earlier symmetry
determinations and stress the uncertainty of the overall picture.
\citet{Han+97} \citep[see also][]{XH2024} find signs of a regular dipolar
magnetic field in the central part of the Milky Way but \citet{W+17} show that
the antisymmetric RM pattern can be due to a local magnetised bubble.  Faraday
rotation measures of magnetic fields in the halos of a sample of edge-on
galaxies \citep{CHANG-ES2024} reveal no preference for clear, simple symmetry,
showing neither a preference for purely quadrupolar nor dipolar field
structures. Thus, it remains unclear if dipolar magnetic structures indeed
occur in galaxies, but our results give some credence to the claims of dipolar
and mixed-parity structures: earlier disc dynamo models were suggesting the
prevalence of quadrupolar magnetic fields.

%------------------------------------------------------
\subsection{The effect of cosmic rays}\label{sec:CM}

We model cosmic rays in a way similar to \citet{DT2022b} and
\citet{Luiz_R_2015a} using a fluid approximation
\citep[e.g,][]{Parker1969,SRLI1985} where the cosmic ray energy density
$\epsilon_\text{cr}$ is governed by
\begin{equation}
    \deriv{\epsilon_\text{cr}}{t} = \mathcal{Q}(z)
-\nabla\cdot(\epsilon_\text{cr}\vec{u}) - p_\text{cr}\nabla\cdot\vec{u}
-\nabla \cdot\vec{F}\,,
\end{equation}
with $\vec{F}$ the cosmic ray flux defined below, $p_\text{cr} =
\epsilon_\text{cr}(\gamma_\text{cr} - 1)$ is the cosmic ray pressure, where
$\gamma_{\text{cr}}=4/3$ is the adiabatic index of the ultrarelativistic gas
\citep{SL1985}.  The cosmic ray pressure is included in the total pressure in
equation~\eqref{eq:momentum}, and $\mathcal{Q}(z)$ is the cosmic ray source of
the form
\begin{equation}
    \mathcal{Q}(z) = Q_{0}\exp(-\lvert z\rvert^2/h_\text{cr}^2)\,.
\end{equation}
The source term is chosen to replicate the injection of cosmic rays into the
ISM by supernovae. A typical supernova (SN) explosion injects about $10^{51}
\erg$ of energy, of which only a few percent comprises cosmic rays
\citep[e.g.,][]{KOG72,Sc02}.
The scale height of the energy injection is set to $h_\text{cr} = 100\p$
\citep{vdBT91} and the rate to $Q_{0} = 9.4 \times 10^{49} \erg \kpc^{-3}
\Myr^{-1}$ \citep{vdBergh90,vdBT91}.

The cosmic ray flux $\vec{F}$ is introduced in a non-Fickian form justified and
discussed by \citet{Snodin+2005},
\begin{equation}
    \tau_\text{cr} \deriv{F_i}{t} = \kappa_{ij}\nabla_j\epsilon_\text{cr}-F_i\,,
\end{equation}
where $\tau_\text{cr}=10 \Myr$ can be identified with the decorrelation time of
the cosmic ray pitch angles, and $\kappa$ is the diffusion tensor,
\begin{equation}
    \kappa_{ij} = \kappa_{\perp}\updelta_{ij}+(\kappa_{\parallel} -\kappa_{\perp})\hat{B}_i\hat{B}_j,
\end{equation}
where a circumflex denotes a unit vector, The parameters that control the
diffusion of cosmic rays $\kappa_\perp= 3.16 \times 10^{25} \, \rm cm^2s^{-1}$
and $\kappa_\parallel= 1.58 \times 10^{28} \, \rm cm^2s^{-1}$ \citep[][and references
therein]{Rodrigues2016,Ryu2003}.

%--------------------------------------------------------------------
\begin{figure}
    \centering
    \includegraphics[width=0.45\textwidth]{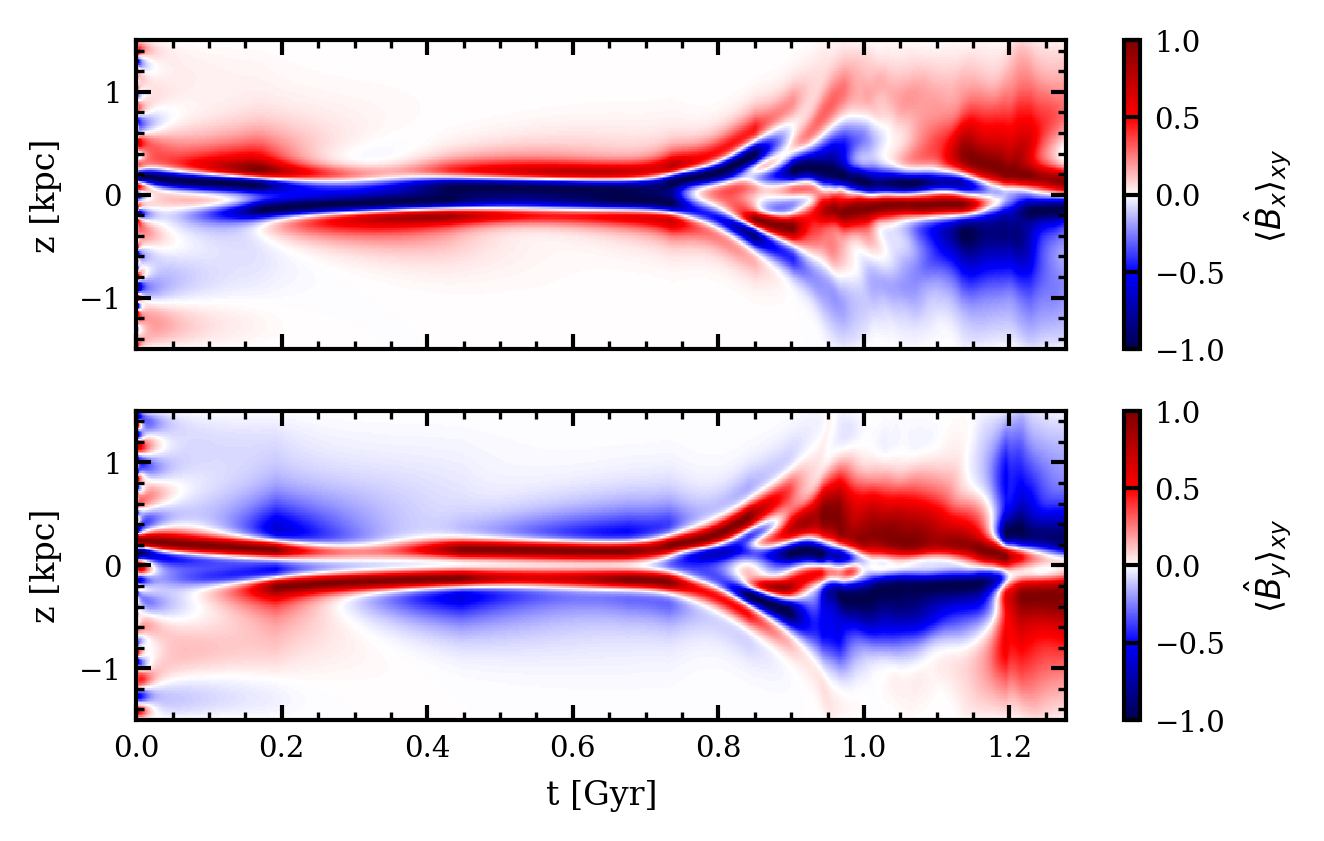}
    \caption{Horizontal averages of the horizontal components of the magnetic
field $\langle \widehat{B}_x \rangle_{xy}$, $\langle \widehat{B}_y
\rangle_{xy}$ in Model~\CM.}
    \label{fig:xy_aver_cm}
\end{figure}
%--------------------------------------------------------------------

Horizontal averages of the magnetic field from Model \CM, which includes cosmic
rays, are shown in Fig.~\ref{fig:xy_aver_cm}, similar to those presented in
Fig.~\ref{fig:xy_averages} for Model \RSOBSD. Cosmic rays enhance the effects
of magnetic buoyancy, thereby strengthening both the MBI and the dynamo action
induced by it. Comparing Models \RSOBSD\ and \CM, the amplification of the MBI
is evident, with the instability growth rate increasing from $\gamma_u = 34.5
\Gyr^{-1}$ to $39.1  \Gyr^{-1}$ due to the inclusion of cosmic rays. Similarly,
the $\alpha^2\Omega$-dynamo growth rate rises from $\gamma_\text{D} = 17.5
\Gyr^{-1}$ to $19.6  \Gyr^{-1}$.

The ratio $\gamma_u/\gamma_\text{D}$ increases when cosmic rays are included,
suggesting that a dipolar non-linear state would be easier to excite in the
presence of cosmic rays.  Models {\RAOASA} and {\RAOASACM} represent the Solar
neighbourhood and the ratio $\gamma_u/\gamma_\text{D}$ increases from 0.83
without cosmic rays to 1.14, while the magnetic field preserves the quadrupolar
parity in the late stages. The $\alpha^2\Omega$-dynamo evolves for $5.5\Gyr$ in
Model {\RAOASA} before the MBI extends the scale height of the quadrupolar
magnetic field. This happens slightly earlier in the models with cosmic rays.
Model {\RBOASACM} uses parameters that represent a galactocentric radius of $R
= 3 \kpc$, with a growth rate $\gamma_\text{D} = 5.4 \Gyr^{-1}$, which is
significantly higher than $1.4 \Gyr^{-1}$ for Model {\RAOASACM}, even though
both have a dynamo number of $D \approx-10$.  It is likely that Model
{\RAOASACM} is sensitive to $\lvert D\rvert$, with $\lvert R_\omega \rvert
\approx 41.6 R_\alpha$, whereas for Model {\RBOASACM}, $\lvert R_\omega \rvert
\simeq 3.9 R_\alpha$, making it more sensitive to the $\alpha$-effect instead.

%--------------------------------------------------
\section{Interpretation of results}\label{sec:interp}

All models follow a similar kinematic evolution characterized by exponential
magnetic energy growth, arising from the combined effects of the imposed
$\alpha$-effect and the $\Omega$-effect due to differential rotation. Once the
magnetic field becomes strong enough to be dynamically significant, the r.m.s.\
velocity grows exponentially. As the kinetic and magnetic energies reach
equipartition, the magnetic field saturates, and the system enters its
non-linear phase, at which point the magnetic field structure changes
substantially. The magnetic field generated by the linear dynamo is
confined to a relatively thin layer, $\lvert z \rvert \lesssim h_{\alpha}$, and
grows monotonically. However, after inducing the MBI, it spreads to larger
altitudes through buoyancy, acquiring a scale height on the order of $1 \kpc$.
When fully non-linear, the magnetic field undergoes a dramatic structural
change, transitioning from an initially quadrupolar field structure to a
dipolar field structure in models \RSOBSB--\RSOBSE. Model {\RSOBSA} has a
higher rate of shear than other models and does not display the same change in
parity, which may be due to differential rotation suppressing the MBI.

To understand the evolution of the system, including the magnetic field parity
variations and what is the role of the MBI, we consider the mean-field
induction equation, written in terms of the mean magnetic field
$\aver{\vec{B}}$  as \citep[see, e.g.,][for details]{SS21}
\begin{equation}\label{MFD}
    \deriv{\aver{\BB}}{t} = \nabla \times \left(\vec{U}\times \aver{\BB}
+\vec{\mathcal{E}} -\eta\nabla\times\aver{\BB}\right)\,,
\end{equation}
where angular brackets denote a suitable averaging, $\vec{U}$ is the mean
velocity field due to the differential rotation and $\vec{\mathcal{E}}$ is the
mean electromotive force (EMF).  The mean magnetic field is obtained by
smoothing the total field $\vec{B}$ with a Gaussian kernel as in
equation~\eqref{GS} with a smoothing length of $\ell=200\p$, so it is the same
as $\vec{B}_\text{B}$ of Section~\ref{sec:Rotation}. We also introduce $\vec{b}
= \vec{B} - \langle \vec{B}\rangle$, the deviation of the total magnetic field
$\vec{B}$ from its mean $\langle \vec{B} \rangle$ which can be identified with
$\vec{B}_\text{D}$. In terms of the velocity field $\vec{u}$, mainly driven by
the MBI and modified by the Lorentz force in the non-linear stages of the
evolution, we have
\begin{align}
    \vec{\mathcal{E}} &\approx \vec{\alpha}\cdot\aver{\BB} -
\vec{\beta}\cdot\left(\nabla \times \aver{\BB}\right)\,,
\label{eq:EMF}
\end{align}
where $\vec{\alpha}$ and $\vec{\beta}$ are second order tensors representing
the induction effect of the mean helicity of the velocity field $\vec{u}$ and
the associated turbulent magnetic diffusivity, respectively.

%---------------------------------------------------------------------
\begin{figure}
    \centering
    \includegraphics[width=0.5\textwidth]{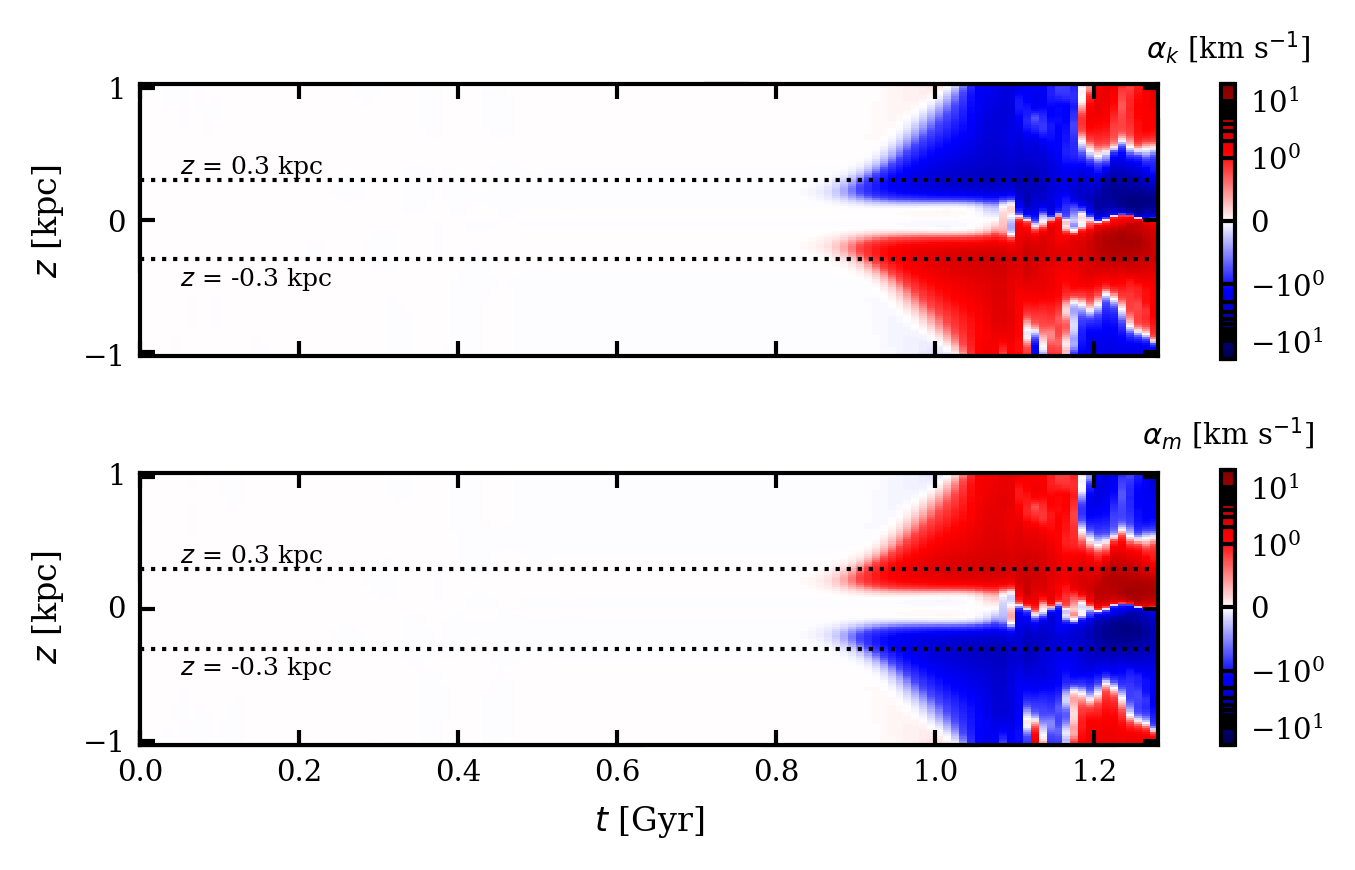}
    \caption{The evolution of the horizontally averaged mean kinetic helicity
(upper panel) and mean magnetic helicity (lower panel),
equations~\eqref{alpha_k} and \eqref{alpha_m}) in Model \RSOBSD. The horizontal
dotted lines are shown at $\lvert z \rvert = h_{\alpha}$.
    }
    \label{fig:ak_am}
\end{figure}
%-----------------------------------------------------------

%---------------------------------------------------------------------
\begin{figure*}
    \centering
    \includegraphics[width=\textwidth]{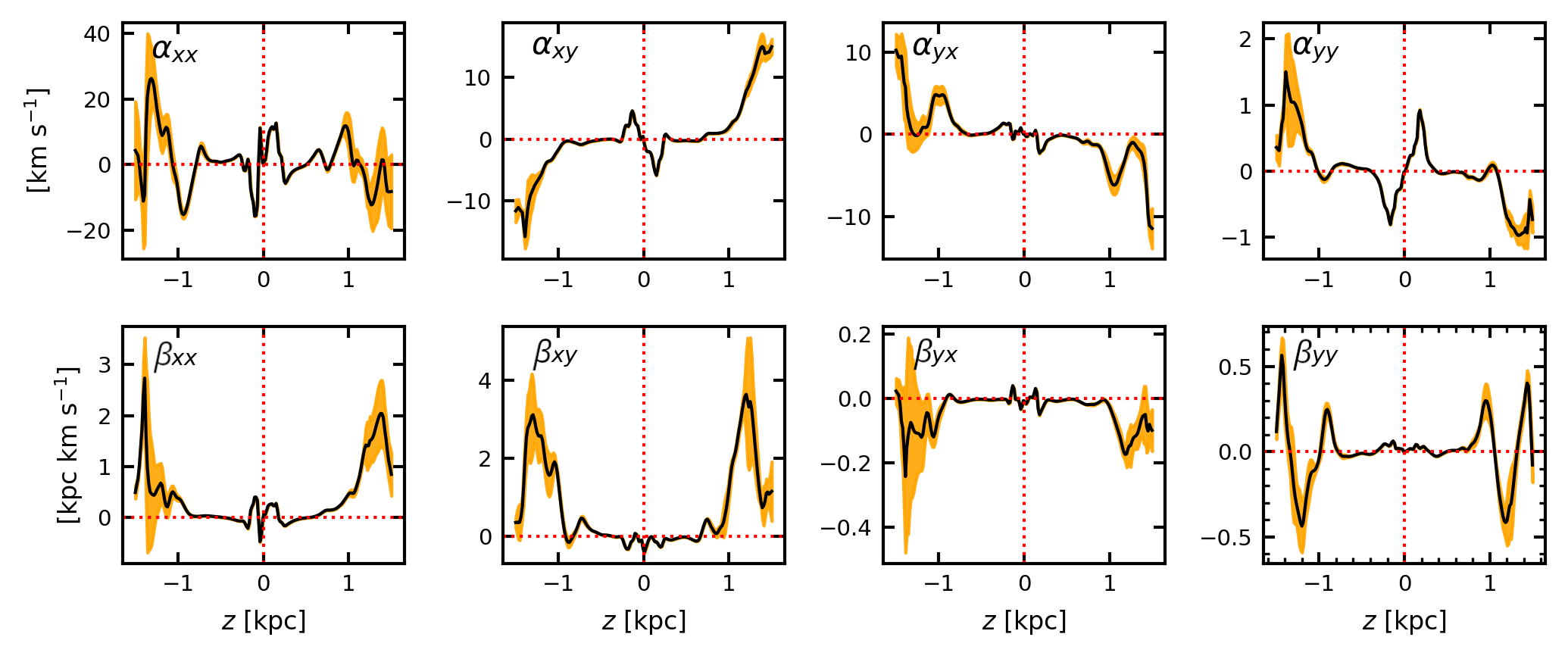}
   \caption{The time-averaged elements of the turbulent transport tensors
introduced in equation~\eqref{EMF} for model \RSOBSD\ {during} the non-linear
state at $1.0<t<1.5 \Gyr$. The yellow shading indicates one standard deviation
of each component based on bootstrap resampling of the time series of {the EMF}
$\vec{\mathcal{E}}$.}
    \label{fig:IROS}
\end{figure*}
%-------------------------------------------------------------------------------

%-----------------------------------------------------------
\subsection{Approximating the $\alpha$-effect}\label{sec:approx}
The Coriolis force acting on gas flows in a stratified disc makes them helical,
with opposite signs of the mean helicity on either side of the midplane.  The
dynamo amplifies a magnetic field, which drives flows with the mean helicity of
the opposite sign, and this quenches the dynamo by reducing the magnitude of
the $\alpha$-effect \citep[Section 7.11 of,][]{SS21}. As a result of the action
of the Lorentz force, the $\alpha$-coefficient is modified as
\begin{equation}
    \alpha=\alpha\kin+\alpha\magn\,,
\end{equation}
where $\alpha\kin$ is due to the mean helicity unaffected by the magnetic field
and $\alpha\magn$ represents the magnetic contribution.

In a simplified case of a scalar helicity coefficients (see
Section~\ref{sec:emf} for a more general analysis), the mean kinetic helicity
coefficient is given by
\begin{equation}\label{alpha_k}
\alpha\kin=-\tfrac13 \tau\mean{\vec{u}\cdot(\nabla\times\vec{u})}\,,
\end{equation}
where $\tau$ is the flow correlation time. The mean helicity produced by the
action of the Coriolis force is such that $\alpha\kin>0$ for $z>0$ and
$\alpha\kin<0$ for $z<0$; this symmetry is adopted in the imposed
$\alpha$-effect of equation~\eqref{eq:alpha}. When the gas velocity field
$\vec{u}$ obtained from simulations presented here is used in this expression,
the result does not include the imposed $\alpha$-effect since it is not
associated with any explicit gas flow. The magnetic part of the mean helicity
is given by
\begin{equation}
    \alpha\magn =
\frac{1}{3}\tau\left\langle\frac{(\nabla\times\vec{b})\cdot\vec{b}}{4\pi\rho}\right\rangle.
    \label{alpha_m}
\end{equation}
{The averaging procedure used here is same as in equation~\eqref{eq:EMF} and both $\alpha\kin$ and $\alpha\magn$ are functions of $x$,$y$ and $z$.}

Figure~\ref{fig:ak_am} presents the horizontally averaged mean kinetic and
magnetic helicity coefficients derived using equations~\eqref{alpha_k} and
\eqref{alpha_m} from $\vec{u}$ and $\vec{b}$ in model \RSOBSD. At $0.75
\lesssim t\lesssim 1.0 \Gyr$, magnetic buoyancy spreads the magnetic field out
of the layer $\lvert z \rvert \lesssim h_{\alpha}$ and both the flow speed and
the Lorentz force become strong enough to make the gas flow significantly
helical.  As expected, both $\alpha\kin$ and $\alpha\magn$ are antisymmetric
with respect to the midplane. However,  the sign of $\alpha\kin$ is opposite to
that produced by the Coriolis force because it is driven by the Lorentz force
of the dynamo-generated magnetic field $\vec{b}=\vec{B}_\text{D}$ which has the
opposite helicity sign. Of course, the Coriolis force also affects the velocity
field $\vec{u}$, and this leads to a change in the sign of $\alpha\kin$ at a
later stage $t\gtrsim1.2\Gyr$ and at large $|z|$ where the magnetic field is
weaker than near the midplane.  The sign of $\alpha\magn$, shown in the lower
panel of Fig.~\ref{fig:ak_am}, is, as expected, opposite to that of
$\alpha\kin$ at a comparable magnitude.

The anomalous sign of the kinetic helicity of the flows produced by the
magnetic buoyancy in the presence of a mean-field dynamo was first noticed by
\citet[][section~6]{DT2022b} and \citet[][section~5]{QSTGB23}. With a
solid-body rotation explored in those papers, this leads to non-linear
oscillations of the magnetic field. Differential rotation drives a stronger
dynamo action and, correspondingly, a stronger MBI explored here. This causes
changes in the large-scale magnetic field polarity.

%----------------------------------------------------------------------------
\subsection{IROS analysis of the EMF composition}\label{sec:emf}

To further verify and justify our interpretation of the results, we have
computed the components of the (pseudo-)tensor $\alpha_{ij}$ and tensor
$\beta_{ij}$ using the method of iterative removal of sources (IROS) introduced
by \citet{bendre2023iterative}. Using sliding time averages of the mean
magnetic field,  the components of the electromotive force $\mathcal{E}_i =
\langle \vec{u}\times\vec{b} \rangle_i$ are approximated by $\mathcal{E}_i = \alpha_{ij}\langle \vec{B_j}\rangle -
\beta_{ij}(\nabla\times \langle \vec{B} \rangle)_j$ 
%$\AB{\xcancel{\mathcal{E}_i =
% (\vec{U}\times\langle\vec{B}\rangle)_i + \alpha_{ij}\langle \vec{B_j}\rangle -
% \beta_{ij}(\nabla\times \langle \vec{B} \rangle)_j}}$. 
Explicitly,
%-----------------------------------------------------------------------------
\begin{align}\label{EMF}
\begin{pmatrix} \mathcal{E}_x\\ \mathcal{E}_y\end{pmatrix}
= \begin{pmatrix} \alpha_{xx} &\alpha_{xy}\\ \alpha_{yx} &\alpha_{yy} \end{pmatrix}
\begin{pmatrix} \mean{B}_x \\ \mean{B}_y \end{pmatrix} 
&-\begin{pmatrix} \beta_{xx} &\beta_{xy}\\ \beta_{yx} &\beta_{yy} \end{pmatrix}
\begin{pmatrix} (\nabla\times\mean{\vec{B}})_x \\ (\nabla\times\mean{\vec{B}})_y \end{pmatrix},
\end{align}
%
% \AB{
% \begin{align}\label{EMF}
% \xcancel{\begin{pmatrix} \mathcal{E}_x\\ \mathcal{E}_y\end{pmatrix}}
% = \xcancel{\begin{pmatrix} (\vec{U}\times\mean{B})_x \\ (\vec{U}\times\mean{B})_y \end{pmatrix}}
% &+ \xcancel{\begin{pmatrix} \alpha_{xx} &\alpha_{xy}\\ \alpha_{yx} &\alpha_{yy} \end{pmatrix}
% \begin{pmatrix} \mean{B}_x \\ \mean{B}_y \end{pmatrix}} \nonumber\\
% &-\xcancel{\begin{pmatrix} \beta_{xx} &\beta_{xy}\\ \beta_{yx} &\beta_{yy} \end{pmatrix}
% \begin{pmatrix} (\nabla\times\mean{\vec{B}})_x \\ (\nabla\times\mean{\vec{B}})_y \end{pmatrix}},
% \end{align}}
%-----------------------------------------------------------------------------
are solved to determine the elements of the tensors $\alpha_{ij}$ and
$\beta_{ij}$, which are assumed to be independent of time. This assumption is
valid in either the early stages of the exponential growth of the magnetic
field or in the later, stationary state of the system.
%\as{In this analysis, we neglect the part of the electromotive force $\vec{U}\times\vec{B}$.}
{These calculations use
the horizontal averaging, $\mean{\vec{B}} = \meanh{\vec{B}}$ as displayed in
Fig.~\ref{fig:xy_averages}, such that the tensor elements are functions of $z$
alone.} The horizontal average of the vertical component of the magnetic field
vanishes due to the horizontal periodic boundary conditions. Hence, the
analysis is applied only to the horizontal components of the magnetic field.

The diagonal elements of the $\alpha$-tensor represent the scalar
$\alpha$-effect discussed in Section~\ref{sec:approx}, with
$\alpha\kin+\alpha\magn \approx (\alpha_{xx}+\alpha_{yy})/2$. If the flow is
isotropic in the ($x,y$)-plane $\alpha_{ij}$ is antisymmetric ($\alpha_{yx} =
-\alpha_{xy}$) and the off-diagonal elements represent the transfer of the mean
magnetic field along the $z$-axis at the effective speed $U_z = -\alpha_{xy}$
due to the increase in the turbulent magnetic diffusivity with $\lvert z
\rvert$ resulting mainly from the increase of the random flow speed
\citep[turbulent diamagnetism -- e.g., Section~7.9 of][]{SS21}. The diagonal
components of the tensor $\beta_{ij}$ represent the turbulent magnetic
diffusion.

Fig.~\ref{fig:IROS} presents the resulting components of the tensors
$\alpha_{ij}$ and $\beta_{ij}$ for the non-linear stage of the evolution. The
yellow shading spans one standard deviation of the variables obtained from five
estimates, each resulting from the sampling of every fifth iteration of 1500,
with intervals of $1\Myr$ in the time series of $\vec{\mathcal{E}}$ at each
$z$.

The sum $\alpha_{xx} + \alpha_{yy}$ is significant in magnitude, antisymmetric
with respect to the midplane $z = 0$, and mostly negative at $z > 0$. The
magnitudes of $\alpha_{xx} + \alpha_{yy}$ are close to $\alpha\kin +
\alpha\magn$ obtained using equations \eqref{alpha_k} and \eqref{alpha_m} at
$t\gtrsim1\Gyr$. The off-diagonal components of $\alpha_{ij}$ are quite close
to the expected antisymmetry, $\alpha_{yx} = -\alpha_{xy}$. Near the midplane,
these support an inward transfer of the mean magnetic field. In association
with the increase in the turbulent magnetic diffusivity with $\lvert z \rvert$,
this will tend to oppose the buoyancy migration of the magnetic field away from
the midplane, thus facilitating the saturation of the MBI.  To confirm our
conclusion that the dynamo action and the associated complex behaviour of the
mean magnetic field are essentially non-linear phenomena, we have verified that
the components of the tensors $\alpha_{ij}$ and $\beta_{ij}$ fluctuate around
the zero level during the linear stage without any significant effect on the
system's evolution.

%-----------------------------------------------------------------
\subsection{One-dimensional mean-field model}\label{sec:1D}

This section presents a non-linear one-dimensional (1D) model designed to
replicate the 3D MHD solutions. We model the mean-field dynamo with advection
due to magnetic buoyancy and show that it not only captures the parity switches
observed in the 3D models but also qualitatively reproduces the resultant
magnetic field.  The Cartesian components of the mean-field dynamo
equation~\eqref{MFD} with $\vec{U}=(U_x,U_y,U_z)$ and scalar $\alpha$ and $\beta$
are written as 
% \fag{If $U_x$=0, why include in these equations or solve Eq 28?}
\begin{align}
\deriv{B_x}{t} &= {k}_y\alpha B_z-\deriv{}{z}(\alpha B_y)+\beta\dfrac{\partial^2 B_x}{\partial z^2}\nonumber \\ &+ {k}_y (U_xB_y - U_yB_x)- \deriv{}{z} (U_zB_x - U_xB_z)\,,\label{1DBx} \\
\deriv{B_y}{t} &= \deriv{}{z}(\alpha B_x)- {k}_x\alpha B_z+\beta\dfrac{\partial^2 B_y}{\partial z^2}\nonumber \\ &+ \deriv{}{z}(U_zB_y - U_yB_z)-{k}_x (U_xB_y - U_yB_x) - SB_x\,,\label{1DBy} \\
\deriv{B_z}{t} &= {k}_x\alpha B_y - {k}_y\alpha B_x  + \beta\dfrac{\partial^2B_z}{\partial z^2}\nonumber \\ &+ {k}_x (U_zB_x - U_xB_z)-{k}_y (U_yB_z - U_zB_y) \,,\label{1DBz}
\end{align}
%\AS{Yasin, in the last equation, the term $\partial(\alpha B_y)/\partial z$ is spurious -- it cannot be there. This is the $z$-component of $\nabla\times(\alpha\vec{B})$, and it can only contain derivatives in $x$ and $y$ but not in $z$. If the 1D model cannot be made working with correct equations, we should just remove it from the paper.}
where $S=-18 \kms\kpc^{-1}$ is the velocity shear rate, and we retain dependence on $t$ and $z$ only (the infinite slab approximation). Here,
$\alpha$ is defined as in equation~\eqref{eq:alpha}, and $\beta = \eta +
\eta_T$ is the sum of the microscopic diffusivity $\eta$ and the turbulent
diffusivity $\eta_T$. {The derivatives $\partial/\partial{x}$ and $\partial{/\partial y}$ are replaced by $k_x=1\kpc^{-1}$ and $k_y=1\kpc^{-1}$, respectively, because $B_x$ and $B_y$ vary in the horizontal direction at approximately these wavenumbers, as evident in Fig.~\ref{fig:field_evolution}.} We use $\beta = 10^{26} \cm^2 \s^{-1}
= \frac13\kpc^{2} \Gyr^{-1}$, which matches the value used in the 3D
simulations (see Table~\ref{tab:sims}). The initial magnetic field {is Gaussian random noise} and has a strength of $10^{-3}\mkG$.

%------------------------------------------------------------------------------
\begin{figure}
    \centering
    \includegraphics[width=0.5\textwidth]{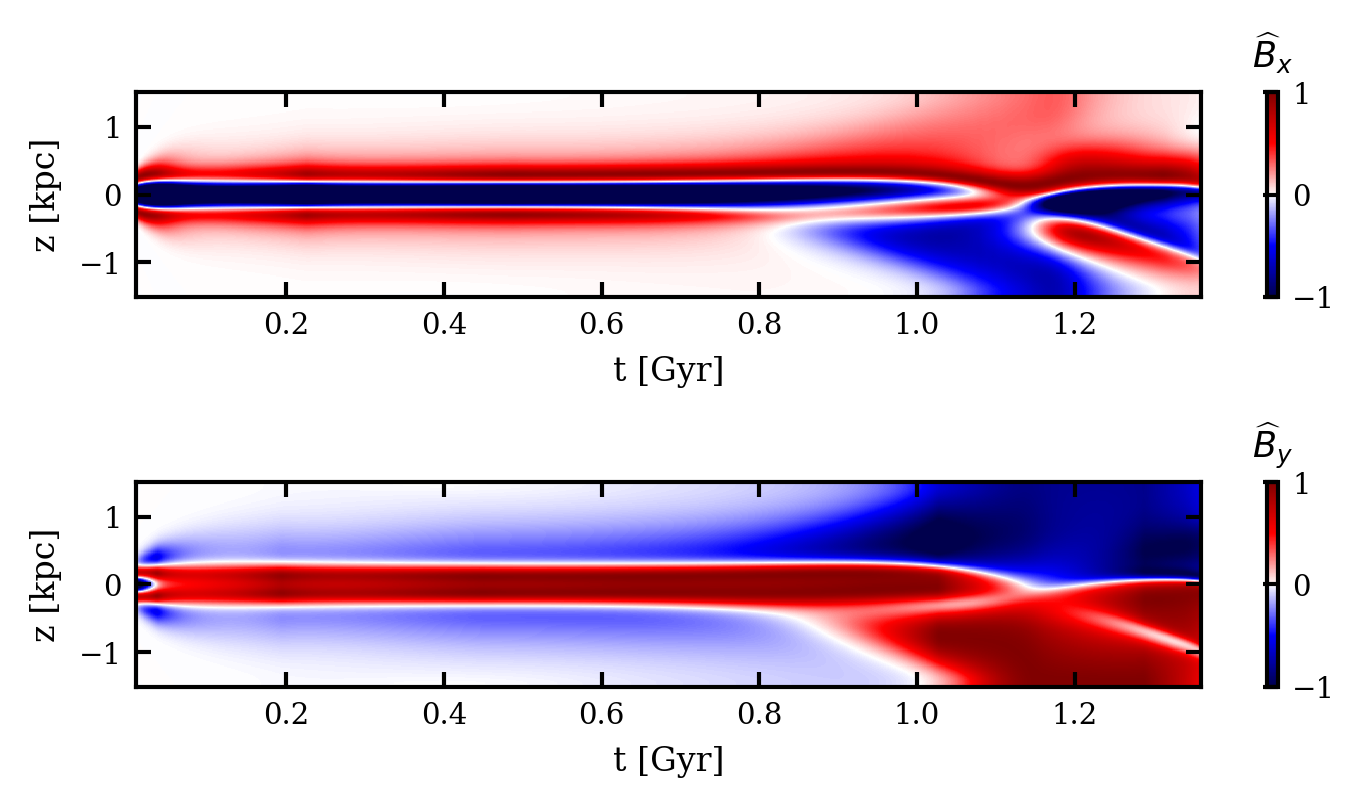}
    \caption{The evolution magnetic field components $\widehat{B}_x$ (upper
panel) and $\widehat{B}_y$ (lower panel) for the 1D model using parameters
which match model {\RSOBSD}. The magnetic field components are normalized to
their maximum values at each time.}
    \label{fig:1D_xy}
\end{figure}
%------------------------------------------------------------------------------
\begin{figure}
    \centering
    \includegraphics[width=0.5\textwidth]{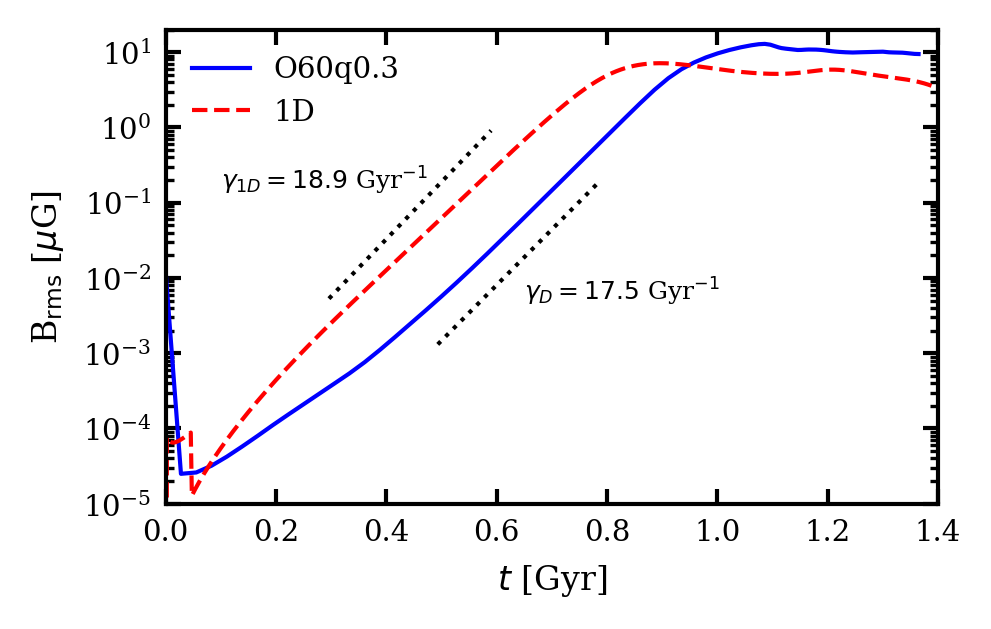}
    \caption{The evolution root mean square magnetic field strength for the  Model {\RSOBSD} (blue solid line) and the 1D model (red dashed line). The dotted black lines are representative of the growth rates of the magnetic field, where $\gamma_{1D}$ is the growth rate of the magnetic field for the 1D model.}
    \label{fig:1D_ts}
\end{figure}
%------------------------------------------------------------------------------
We omit brackets denoting the averaging to simplify notation in this section,
including the velocities $U_x$, $U_y$ and $U_z$
% \fag{To clarify, $U_x=0$?}
which are understood as averages over an intermediate scale between the scale
of the random motions produced by the MBI and the vertical extent of the system
$|z|\leq z_0$.  Together with the equations shown above, we include the
Cartesian component of the Navier--Stokes equation,
\begin{align}
    \deriv{U_x}{t} &= \dfrac{1}{4\pi\rho_0}B_z\deriv{B_x}{z} + 2U_y\Omega + \nu \frac{\partial^2 U_x}{\partial z^2}\label{1DUx}\,,\\
    \deriv{U_y}{t} &= \dfrac{1}{4\pi\rho_0}B_z\deriv{B_y}{z} - 2U_x\Omega - SU_x  + \nu \frac{\partial^2 U_y}{\partial z^2} \label{1DUy}\,,\\
    \deriv{U_z}{t} &= - \dfrac{1}{8\pi\rho_0}\deriv{}{z}\left({{B_x}^2 +{B_y}^2} \right) +\frac{\rho^\prime}{\rho_0}g + \nu \frac{\partial^2 U_z}{\partial z^2}  \,,\label{1DUz}
\end{align}
%------------------------------------------------------------------------------
% \fag{Yasin, please confirm that these equations are now correct. Note the pressure and tension force have opposite sign.}
% % \AS{Yasin, where from the first two terms on the rhs of eq.~\eqref{1DUz} come? They look like a magnetic pressure gradient, albeit without the factor $1/4\pi$, but this is included in the second line of that equation. Please check and correct either the text or the model.}
where $g$ is the vertical acceleration due to gravity
% , magnetic tension \fag{Magnetic tension is the first term in each equation?} is neglected
and the initial velocity is set to zero. We neglect the time and space variations of
the gas density, adopting $\rho = \rho_0$ at all times
% \AS{Why then $\nabla\rho_0$ is included?}
but, in the spirit of the Boussinesq approximation, include the density
variation $\rho^\prime$ in the Archimedes force. Consider a region with density
$\rho = \rho_0 + \rho^\prime$, containing a magnetic field $\vec{B}+\vec{b}$
and surrounded by gas of density $\rho_0$ with a magnetic field {$\vec{B}$}.
Here $\vec{B}$ is the mean field and $\vec{b}$ is its local (in $z$)
perturbation, which is calculated by smoothing $\vec{B}$ in $z$ using the Gaussian filter and subtracting it from the total
field. The smoothing length used is the same as in the 3D simulations,
$\ell=200\p$. The pressure balance in an isothermal gas then leads to
%------------------------------------------------------------------------------
\begin{equation}
\rho^\prime = -\dfrac{2Bb + b^2}{8\pi c_\text{s}^2}\,.
\end{equation}
% \AS{Yasin, how $\vec{b}$ is obtained in the 1D model? You say you are using Gaussian smoothing in $z$ at each time step? Is this indeed true? If so, what is the separation scale?  Or did you use $\vec{b}$ obtained from a 3D solution? Then the 1D model is not self-consistent/closed.}

Given this 1D model applies only along $z$, $\nabla \cdot \mathbf{B} = 0$
implies that $\partial\langle B_z \rangle / \partial z = 0$,  as horizontal
derivatives of $B_x$ and $B_y$ must vanish. However, the 1D model presented
here fails to reproduce the magnetic parity changes if $\partial B_z/\partial
z=0$ is assumed for the mean field.  Only if we permit averaging to be a local
property, which permits deviations in the plane and hence local derivatives of
$B_z$, as with the use of Gaussian smoothing, such that $\nabla\cdot\vec{B}=0$
is satisfied locally. Thus, important information is lost when considering
horizontally averaged fields and the horizontal spatial structure of all
components of the magnetic field should be accounted for explicitly, as in the
3D simulations, or implicitly, as in the 1D model.
%\ass{Crucially, the averaging condition used} \as{The averages considered in
%this section} must exclude the horizontal average, which is applied in
%Section~\ref{sec:emf} and Figs.~\ref{fig:xy_averages}--\ref{fig:IROS} \as{for
%the presentation purposes only}, since the field averaged in the horizontal
%$(x,y)$-planes is a function of $z$ alone. \ass{Combined with $\nabla \cdot
%\mathbf{B} = 0 \Rightarrow \partial\langle B_z \rangle / \partial z = 0$.}
%\as{Then, in 1D, $\nabla\cdot\vec{B}=0$ implies $\partial B_z/\partial z=0$
%for the horizontal averages.} \ass{However, its restrictions limit the
%admissible structure of the magnetic field, without any physical or
%mathematical justification. If such an averaging method were used, there could
%be no evolution equation for $B_z$. This indicates that} \as{Thus,} important
%information is lost when considering \ass{a} horizontally averaged field\as{s}
%and\ass{ that} the horizontal spatial structure of all components of the
%magnetic field should be accounted for \as{explicitly, as in the 3D
%simulations, or implicitly, as in the 1D model}. \ass{as they may produce
%non-trivial effects.} \as{In particular, the 1D model presented here fails to
%reproduce the magnetic parity changes if $\partial B_z/\partial z=0$ is
%assumed for the mean field.}

Equations \eqref{1DBx}--\eqref{1DUz} are solved numerically in $-z_0 < z < z_0$
with $z_0 = 1.5 \kpc$. The seed magnetic field is a Gaussian random noise, and the imposed
$\alpha$-effect generates a quadrupolar magnetic field.  At $z=\pm z_0$, we
apply an impenetrable boundary condition for $U_z$, vacuum boundary conditions
for the magnetic field, and assume, for simplicity, that $B_z$ also vanishes
%------------------------------------------------------------------------------
\begin{equation}
    U_z = B_x = B_y = B_z = 0\,.
\end{equation}
%------------------------------------------------------------------------------
Larger vertical sizes $z_0$ were tested to confirm that the domain is large
enough to prevent spurious boundary effects.

Figure~\ref{fig:1D_xy} shows the evolution of the horizontally averaged magnetic
field components in the 1D model, including the change in the magnetic field
parity from quadrupolar to dipolar.  It qualitatively reproduces the evolution
shown in Fig.~\ref{fig:xy_averages}, including the time scale of the parity
reversal. We do not attempt to achieve a precise match between the {3D} and
{1D} results, being content with the fact that the {1D} model further justifies
our conclusion that the change in field parity is a non-linear phenomenon that
relies on the interaction of the mean-field dynamo enhanced by the differential
rotation and magnetic buoyancy.
{Figure~\ref{fig:1D_ts} compares {the evolution of} the root-mean-square strengths of the magnetic {field in model} {\RSOBSD} (solid blue line) and the one-dimensional model (dashed red line). {The growth rates are quite similar in the 3D simulations and the 1D model, $\gamma_D=17.5\Gyr^{-1}$ and $\gamma_{1D}=18.9\Gyr^{-1}$, respectively.} However the one-dimensional model saturates at a {somewhat} lower level {of the magnetic field strength}.}
The key parameters that influence the solution are the vertical extent
$h_\alpha$ of the imposed $\alpha$-effect, the value of $\alpha_0$, the shear
rate $S$, and the diffusion coefficients $\nu$ and $\beta$.  Reducing the shear
rate $S$, which decreases the dynamo number $D$ for a given $h_\alpha$,
enhances the relative strength of the MBI and facilitates the transition to a
dipolar non-linear state. Increasing $\alpha_0$ strengthens the $\alpha^2$
dynamo, promoting quadrupolar solutions in both the linear and non-linear
phases. Conversely, decreasing $\alpha_0$ shifts dominance to the
$\alpha\Omega$-dynamo, encouraging a dipolar field structure in the non-linear
phase -- an effect similarly achieved by increasing $S$.  Increasing $h_\alpha$
favours dipolar magnetic configurations. The viscosity and magnetic diffusivity
are set equal, $\nu = \beta = 0.3  \mathrm{kpc}^2  \mathrm{Gyr}^{-1}$ and
increasing either parameter reduces the dynamo growth rate. If both are
sufficiently large, both the dynamo and the MBI can be completely suppressed.

%----------------------------------------------------------
\section{Summary and implications}\label{sec:Summary}

The non-linear interaction between the mean-field dynamo and magnetic buoyancy
leads to profound changes in the evolution of the large-scale magnetic field.
Magnetic buoyancy spreads the magnetic field into the corona of the galaxy. The
helical large-scale magnetic field which originates from the mean-field dynamo
action in the disc produces gas flows with a kinetic helicity of the opposite
sign to the helicity produced by the Coriolis force and associated with the
$\alpha$-effect near the midplane from which the large-scale field has been
generated.  These flows at large $|z|$ drive a secondary mean-field dynamo
action but, since the sign of their $\alpha$-effect is opposite to that in the
early (kinematic) stages of the dynamo, the resulting magnetic field can have a
dipolar parity, opposite to that of the kinematic dynamo in a thin disc. In
addition, the system develops non-linear oscillations of the magnetic field
which also do not occur in a kinematic disc dynamo.

When the buoyant magnetic field is not helical (e.g., unidirectional) and there
is no rotation, magnetic buoyancy will only redistribute the large-scale
magnetic field to higher altitudes, significantly reducing its pressure
gradient. This leaves the support of the gas layer against gravity to rely
solely on the thermal pressure gradient, along with contributions from
turbulence and random magnetic fields, if present \citep{DT2022b}.

The inclusion of rotation changes the picture because the gas flows that
accompany magnetic buoyancy become helical, driving a mean-field dynamo. As
shown in \citet{DT2022a}, this dynamo can overwhelm an imposed magnetic field,
leading to a reversal, which suggests the potential for magnetic oscillations.
In our case, the resultant complex structure of the kinetic helicity, which
arises from the combined effects of the Lorentz and Coriolis forces, is
responsible for the change in magnetic field parity observed in the non-linear
stages of the system. \citet{DT2022a} examine the effects of rotation and shear
on an imposed magnetic field, rather than a dynamo-generated field, but do not
obtain dipolar magnetic fields. This emphasizes the importance of the
consistent inclusion of the turbulent dynamo action in all parts of the system.

The consequences of the interactions between the turbulent dynamo and the
magnetic buoyancy instability depend on the relative intensity of each process.
This will vary both across different locations within a galaxy and between
different galaxies. The intensity of the dynamo action increases with the scale
height of the gas and with higher velocity shear due to differential rotation.
Meanwhile, the efficiency of the magnetic buoyancy instability is enhanced as
the scale height of the horizontal magnetic field reduces.  The outcome of
their interaction therefore depends primarily on the values of the disc
thickness and the strength of the differential rotation. In general, the
effects of the MBI would be most apparent when the disc is particularly thin
and supports a strong planar magnetic field.  Such conditions are likely to
occur within the inner few kiloparsecs of a spiral galaxy.  Differential
rotation increases the strength of both the $\alpha^2\Omega$-dynamo and the
magnetic buoyancy instability (MBI) compared to similar models in
\citet{QSTGB23} where a solid-body rotation is considered.

Even-parity large-scale magnetic fields in galactic discs have been a firm
prediction of the versions of the galactic dynamo theory \citep{SS21}
where the non-linear interaction with the magnetic buoyancy is not included.
Our results show that a
dipolar magnetic field can be maintained within
$1\text{--}3\kpc$ of the galactic
centre and, depending on the parameters, also in the outer parts of galaxies.
We note, however, that the model presented here neglects the variation of
the angular velocity $\Omega$ with $|z|$. This is justifiable within the region
considered, $|z|\lesssim1.5\kpc$ (see Section~\ref{sec:Equations}), but a
substantial decrease in $\Omega$ with $|z|$ would reduce the intensity of the
secondary dynamo action in the corona. The consequences of such a decrease
remain to be explored.

The simulations presented in this work are conducted in a relatively large but
finite part of the gas layer ($2 \times 2\kpc^2$ horizontally) using
Cartesian coordinates. The computational domain is sufficiently large to
accommodate the most rapidly growing mode of the MBI, and the results are
unlikely to differ significantly in cylindrical coordinates, where the unstable
magnetic field is not strictly unidirectional.
%FAG: swapped as for yq as this relates to response to referee
 {We have confirmed that the large-scale magnetic tension force expected to
arise in cylindrical coordinates, $|(\vec{B} \cdot \nabla) \vec{B}|/(8\pi\rho)
\simeq B^2/(8\pi\rho R)$ where $R\simeq10\kpc$ is the assumed distance to the
galactic centre, is much smaller than the other relevant forces,
$|2\vec{\Omega} \times \vec{u}|$ and $|(\vec{u} \cdot \nabla) \vec{u}|$.}
Thus, our main conclusions should be applicable to disc galaxies and accretion
discs in general, at least at some distance from the disc axis where curvature
effects are weaker.
{\citet{STS2019} and
\citet{S2023} use  cylindrical shearing box
simulations to
investigate the magneto-rotational instability (MRI) and compare
the results with those obtained in a Cartesian
shearing box. These authors find that these approaches
lead to similar results and note that the main advantage
of the cylindrical shearing box approach is the computational efficiency.
%FAG: I have resolved Anvar's revisions into this paragraph, please confirm
%FAG: this is as intended.
% \AS{This argument is not acceptable since the magnetic field in our
% simulations is unidirectional, so that the magnetic tension must be much
% smaller than the other forces by construction. A proper estimate would
% compare an expected tension force in curved magnetic field with the other
% forces in our model. I have reformulated this in the previous paragraph --
% please confirm if what I've written is correct.}
}

\section*{Acknowledgements}
The authors benefited from valuable discussions at the Nordita workshop
`Towards a Comprehensive Model of the Galactic Magnetic Field' at Nordita
(Stockholm) in 2023, supported by NordForsk and Royal Astronomical Society.
FAG acknowledges support of the Finnish Ministry of Education and Culture
Global Programme USA Pilot 9758121 and the Swedish Research Council
(Vetenskapsrådet) grant No.~2022– 03767. Ahbijit B.~Bendre acknowledges funding
from the Italian Ministry for Universities and Research (MUR) through the
"Young Researchers" funding call (Project MSCA 000074).

%%%%%%%%%%%%%%%%%%%%%%%%%%%%%%%%%%%%%%%%%%%%%%%%%%
\section*{Data Availability}

The raw data for this work were obtained from numerical simulations using the
open source PENCIL-CODE available at
\url{https://github.com/pencil-code/pencil-code.git}. The derived data used for
the analysis are available from the corresponding author on a reasonable
request.

% \AS{\bf Yasin, please check carefully the reference. In particular, make sure that the standard MNRAS journal title abbreviations are used throughout -- it is easy to see that both ApJ and Astrophysical Journal are shown, as well as A\&A and Astronomy and Astrophysics -- please check ALL unabbreviated journal titles -- e.g., Reviews of Modern Physics, Geophysical \& Astrophysical Fluid Dynamics, Astronomische Nachrichten.}



\bibliographystyle{mnras}
\bibliography{refs} % if your bibtex file is called example.bib

%\begin{thebibliography}{99}
%\end{thebibliography}

%%%%%%%%%%%%%%%%% APPENDICES %%%%%%%%%%%%%%%%%%%%%

% \appendix

% \section{Some extra material}

% If you want to present additional material which would interrupt the flow of the main paper,
% it can be placed in an Appendix which appears after the list of references.

%%%%%%%%%%%%%%%%%%%%%%%%%%%%%%%%%%%%%%%%%%%%%%%%%%

% Don't change these lines
\bsp	% typesetting comment
\label{lastpage}
\end{document}